\documentclass[final,5p,times,twocolumn]{elsarticle}

\usepackage{nicefrac}

\usepackage{amssymb}
\usepackage{amsthm}
\usepackage{amsmath}
\usepackage{siunitx}
\usepackage{makecell}
\graphicspath{{images}}
\usepackage{subcaption}

\journal{Acta Astronautica}

\begin{document}

\begin{frontmatter}

\title{A planning tool for optimal three-dimensional formation flight maneuvers of satellites in VLEO using aerodynamic lift and drag via yaw angle deviations}

\author[inst1]{Constantin Traub}
\ead{ctraub@irs.uni-stuttgart.de}

\address[inst1]{Institute of Space Systems (IRS), University of Stuttgart, Pfaffenwaldring 29, 70569 Stuttgart, Germany}

\author[inst1]{Stefanos Fasoulas}
\author[inst1]{Georg H. Herdrich}

\begin{abstract}
Differential drag is a promising option to control the relative motion of distributed satellites in the Very Low Earth Orbit regime which are not equipped with dedicated thrusting devices. A major downside of the methodology, however, is that its control authority is (mainly) limited to the in-plane relative motion control. By additionally applying differential lift, however, all three translational degrees-of-freedom become controllable. In this article, we present a tool to flexibly plan optimal three-dimensional formation flight maneuvers via differential lift and drag. In the planning process, the most significant perturbing effects in this orbital regime, namely the $ J_{2} $ effect and atmospheric forces, are taken into account. Moreover, varying atmospheric densities as well as the co-rotation of the atmosphere are considered. Besides its flexible and high-fidelity nature, the major assets of the proposed methodology are that the in-and out-of-plane relative motion are controlled simultaneously via deviations in the yaw angles of the respective satellites and that the planned trajectory is optimal in a sense that the overall decay during the maneuver is minimized. Thereby, the remaining lifetime of the satellites is maximized and the practicability and sustainability of the methodology significantly increased. To the best of the authors knowledge, a tool with the given capabilities has not yet been presented in literature. The resulting trajectories for three fundamentally different relevant formation flight maneuvers are presented and discussed in detail in order to indicate the vast range of applicability of the tool.
\end{abstract}


\end{frontmatter}

\section{Introduction}
\label{Introduction}
Due to benefits such as enhanced redundancy, flexibility and because it renders new scientific methods possible, there is a recent tendency to replace bulky monolithic satellites by several small, distributed satellites flying in formation. During the operational lifetime of a formation, different mission phases may require different formation geometries and a proper reconfiguration strategy is indispensable. To this day, the required control force is generated by means of chemical and/or electric thrusters. However, their utilization has detrimental effects on small satellites’ limited mass, volume and power budgets so that thruster-less alternatives are of highest interest to the small-satellite community. According to best system engineering practice, it is the major disturbance effect which should be exploited for actuation purpose. In Very Low Earth Orbit (VLEO), defined as the entirety of orbits with a mean altitude lower than 450 km \cite{Crisp.2020}, these are the Earth’s oblateness and the aerodynamic drag. Consequently, in the mid-eighties Leonard introduced differential drag as a promising option for the propellant-less control of satellite formation flight \cite{Leonard.1986,Leonard.1989}. The method consists of intentionally creating differences in the magnitudes of aerodynamic drag experienced by two or more spacecraft flying in formation. Since then, it has been continuously studied and even successfully demonstrated in-orbit. A comprehensive literature review of the research field can be found in \cite{Traub.2019}. Notably, due to the sheer increase in the number of satellites planned in the LEO regime within the next ten years \cite{Alfano.2020}, the ability to actively perform collision avoidance maneuvers will become vital for any satellite orbiting in this regime. Thus, especially for satellites which are not equipped with a dedicated thrusting device, an increase in the relevance of aerodynamic orbit control is expected.\\

At the Institute of Space Systems (IRS) of the University of Stuttgart, this methodology is investigated. To circumvent the inevitable trade-off between the achievable accuracy and the computational burden, two different approaches are followed in parallel \cite{Traub.2020b}. In the first approach, simplified maneuver algorithms are developed, enhanced and refined \cite{Walther.2020,Buhler.2021}. As they are computationally very inexpensive, powerful tools to gain deep insights and to derive general conclusions can be created by applying Monte Carlo (MC) methods. However, these algorithms are based on the assumption of a constant residual atmospheric density and the resulting control is of bang-bang nature, both of which renders them impractical to be implemented in a real mission scenario. Therefore, the second approach comprises the development of high-fidelity maneuver sequences for in- and out-of-plane control using differential drag and lift, respectively. First achievements based on Lyapunov principles, which were based on the excellent developments and findings of D. Pérez and R. Bevilacqua \cite{Perez.2012,PerezPhD.2013}, have been presented in a previous publication, in which additionally the significant influence of satellite surface materials on the achievable differential lift forces and therefore the overall maneuver outcome has been indicated \cite{Traub.2019b}. However, the developed approach only allowed to control either the in-plane relative motion via differential drag or the out-of-plane relative motion via differential lift. Both differential forces, however, could not be exploited simultaneously. In addition, the methodology was rather inflexible, restricted to rendezvous maneuvers and resulted in a bang-bang type control profile. In more recent efforts, a more flexible and enhanced option to plan in-plane formation flight maneuvers of two cooperative satellites using differential drag via pitch angle deviations was developed. With the resulting optimal control approach, which is based on the original approach developed by L. Dell’Elce et al. \cite{DellElce.2015,DellElce.2015b}, studies on the interdependencies of relevant parameters have been conducted \cite{Traub.2020}.\\

In this article, which builds upon several previous articles, an extended version of the optimal control approach, which now evolved into a standalone tool, is presented. The major upgrade is that, by commanding not pitch but yaw angle deviations, differential aerodynamic forces in- and normal to the orbital plane can be exerted simultaneously. As a consequence, the controllability of the control method is extended from two dimensions (in-plane relative motion) to three dimensions (in- and out-of-plane relative motion) so that it can be applied to arbitrary formation flight maneuvers. In addition, the fidelity of the planning process is increased by incorporating more evolved effects such as the co-rotation of the atmosphere with the Earth. As the control option is particularly suited for satellites which are not equipped with dedicated thrusting devices, the loss in specific mechanical energy, which is irreversible and shortens the overall mission lifetime, is minimized by the planner. Thus, the maneuvers are designed while simultaneously maximizing the remaining lifetime of the satellites and thereby increasing the practicability and sustainability of the methodology. \\

The remaining article is structured as follows: in the following chapter, chapter \ref{sec:Background}, the background including the fundamentals of satellite aerodynamics, the methodology of differential lift and drag, the employed coordinate systems, nearly-nonsingular orbital elements as well as the concept of differential orbital elements is introduced and discussed. In chapter \ref{sec:PlanningTool}, the maneuver planning tool is explained in detail. In chapter \ref{sec:Trajectories}, the results of three different example test cases are shown and discussed before in chapter \ref{sec:Conclusion}, conclusions are drawn and future work is discussed. \\

Throughout this article, scalars are indicated in roman letters ($ x $), vectors in bold letters ($ \boldsymbol{x} $), unit vectors in bold letters with a surface hat ($ \hat{\boldsymbol{x}} $) and matrices in bold capital letters which are bracketed in squared brackets ($ \left[ \boldsymbol{X} \right]  $). Mean counterparts of orbital elements are indicated via a superscript bar (Keplerian elements $ \boldsymbol{\mathcal{E}} $ as $ \boldsymbol{\bar{\mathcal{E}}} $ and nearly-nonsingular elements $ \boldsymbol{\mathcal{E}}_{ns} $ as  $ \boldsymbol{\bar{\mathcal{E}}}_{ns} $, respectively). Mean elements are calculated from their osculating counterparts by means of a Brouwer-Lyddane contract transformation \cite{Schaub.2018}.

\section{Background}
\label{sec:Background}

\subsection{Satellite aerodynamics}
\label{ssec:SatAero}

\subsubsection{The fundamentals of aerodynamic lift and drag}
\label{sssec:SatAeroFund}
\begin{figure}
	\centering
	\includegraphics[width=0.75\linewidth]{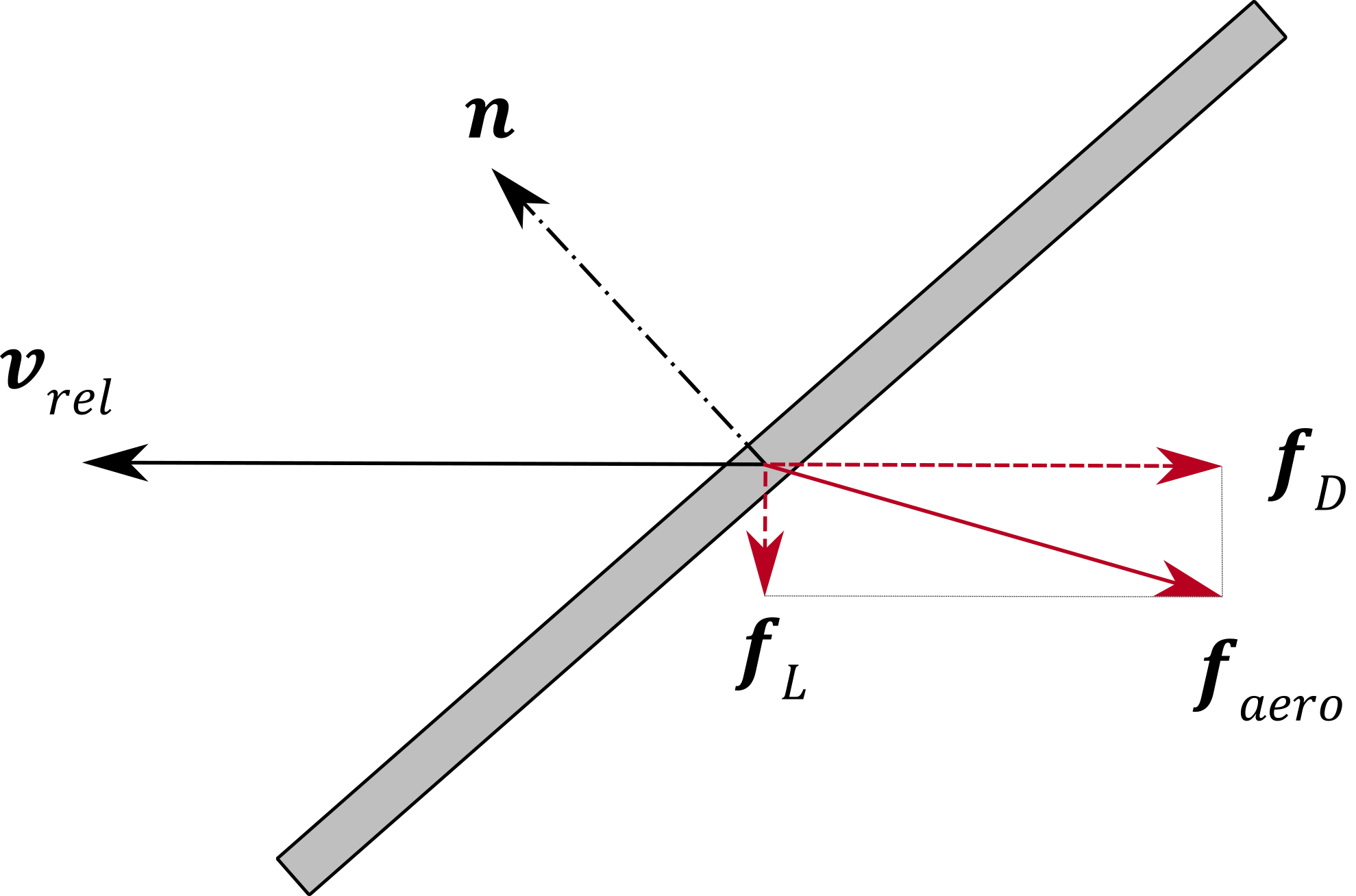}
	\caption{Visualization of the specific drag $ \boldsymbol{f}_{D} $ and lift $ \boldsymbol{f}_{L} $ force as well as the resulting overall aerodynamic force $ \boldsymbol{f}_{aero} $ acting on a simple one sided flat plate with surface normal vector $ \boldsymbol{n} $.}
	\label{fig:FlatPlate}
\end{figure}

Aerodynamic drag, a non-conservative perturbation force that is a result of the interchange of momentum between the Earth’s atmosphere and the spacecraft surface, retards the motion of satellites orbiting in VLEO. Following the description of Vallado \cite{Vallado.op.2013}, the aerodynamic drag acting on a satellite can be expressed as a specific force $ \boldsymbol{f}_{D} $ as:\\
\begin{equation} \label{eq:Drag}
	\boldsymbol{f}_{D}=-\frac{1}{2} \ \rho \ \frac{_{C_{D}A}}{m}|\boldsymbol{v}_{rel}|^{2} \frac{\boldsymbol{v}_{rel}}{|\boldsymbol{v}_{rel}|}
\end{equation}
which has units of acceleration and is the inertial force per unit mass required to produce, in an inertial reference frame and following from Newton’s second law, the acceleration $ \boldsymbol{a}_{D}$, which is the acceleration of a mass $ m $ proportional to the inertial drag force $ \boldsymbol{F}_{D} $. In Eq. \ref{eq:Drag}, $ \rho $ is the local atmospheric density, $ C_{D} $ the drag coefficient of the spacecraft, $ m $ its mass and $ A $ its cross-sectional area perpendicular to the relative velocity vector $ \boldsymbol{v}_{rel} $, which is measured relative to the local atmosphere taking the atmospheric co-rotation with the Earth and thermospheric winds into account:\\
\begin{equation} \label{eq:RelVel}
	\boldsymbol{v}_{rel} = \boldsymbol{v}_{sat} - \boldsymbol{\omega}_{e} \times \boldsymbol{r}_{sat} - \boldsymbol{v}_{wind}
\end{equation}
In Eq. \ref{eq:RelVel}, $ \boldsymbol{r}_{sat} $ and $ \boldsymbol{v}_{sat} $ are the inertial satellite position and velocity, $ \boldsymbol{\omega}_{e} $ the rotational velocity of the Earth and $ \boldsymbol{v}_{wind} $ is the thermospheric wind velocity. The direction of the specific atmospheric drag force $ \boldsymbol{f}_{D} $ acting on a simple on-sided flat plate is visualized in Fig. \ref{fig:FlatPlate}.

All parameters from Eq. \ref{eq:Drag} which depend on the spacecraft design are commonly combined in the ballistic coefficient $ \beta$:\\
\begin{equation} \label{eq:BalCoef}
 \beta = \frac{m}{C_{D}A} 
\end{equation}
which is a measure of the sensitivity of the spacecraft to perturbations by the drag force. All remaining parameters, which are independent from the spacecraft design, represent the dynamic pressure $ q $:\\
\begin{equation} \label{eq:DynPress}
	 q = \frac{1}{2}\rho |\boldsymbol{v}_{rel}|^{2} 
\end{equation}
In the case of aerodynamic lift, which acts perpendicular to drag, the direction depends on the orientation of the surface under consideration and the specific force $ \boldsymbol{f}_{L} $ can be calculated as:\\
\begin{equation} \label{eq:Lift}
	\boldsymbol{f}_{L}=-\frac{1}{2} \ \rho \ \frac{_{C_{L}A}}{m}|\boldsymbol{v}_{rel}|^{2} \hat{\boldsymbol{u}}_{L}
\end{equation}
with the effective direction $ \hat{\boldsymbol{u}}_{L} $ being defined as:\\
\begin{equation} \label{eq:LiftDir}
	\hat{\boldsymbol{u}}_{L} = \frac{\left(\boldsymbol{v}_{rel} \times \boldsymbol{n}\right) \times  \boldsymbol{v}_{rel}}{|\left(\boldsymbol{v}_{rel} \times \boldsymbol{n}\right) \times  \boldsymbol{v}_{rel}|}
\end{equation}
In Eq. \ref{eq:Lift} and \ref{eq:LiftDir}, $ C_{L} $ is the lift coefficient and $ \boldsymbol{n} $ is the surface normal vector of the surface under consideration. The direction of the specific atmospheric lift force $ \boldsymbol{f}_{L} $ acting on a simple on-sided flat plate with surface normal vector $ \boldsymbol{n} $ is exemplarily visualized in Fig. \ref{fig:FlatPlate}. 

In analogy to drag, all terms which depend on the spacecraft design can be summarized in a parameter which will be referred to as the ballistic lift coefficient $ \beta_{L} $ in the following:\\
\begin{equation} \label{eq:BalCoefLift}
	\beta_{L} = \frac{m}{C_{L}A} 
\end{equation}

In relevant literature, aerodynamic lift has so far predominantly considered to be negligible. This is due to a number of reasons, including that:
\begin{itemize}
\item satellites that are spinning/tumbling tend to have the effect of aerodynamic lift cancel out;
\item satellites with symmetrical shapes, e.g. spherical satellites such as Sputnik, do not produce lift at all; 
\item the lift coefficients $ C_{L} $ experienced in-orbit so far are significantly smaller than the drag coefficients $ C_{D} $ \cite{Horsley.2011b};
\item perpendicular force components are significantly less effective in changing the orbit geometry compared to the along-track force component (according to the perturbation equations) 
\end{itemize}
However, by intentionally maintaining a constant angle-of-attack of relevant surfaces with respect to the relative velocity vector, the effects of aerodynamic lift are shown to essentially build up over time and generate measurable effects on the satellite orbit. This was first experienced during the analysis of the inclination of the S3-1 satellite in 1977 \cite{Ching.1976}. Moore studied the effects of aerodynamic lift on near circular satellite orbits in closer detail in 1985 \cite{Moore.1985}. \\

In conclusion, the aerodynamic force $ \boldsymbol{f}_{aero} $ acting on a satellite can be calculated simply as the sum of both individual forces (which is exemplarily visualized in Fig. \ref{fig:FlatPlate}):\\
\begin{equation} \label{eq:Aero}
	\boldsymbol{f}_{aero}= \boldsymbol{f}_{D} + \boldsymbol{f}_{L}
\end{equation}

\subsubsection{The methodology of differential lift and drag}
\label{sssec:DiffLiftDrag}
Assuming Keplerian orbits, a formation consisting of a reference spacecraft, the so-called chief, and a second spacecraft, from now on referred to as deputy, is long-term stable if and only if the semi-major axes (and therefore the specific mechanical energies) of both satellites are equal. In reality though, any orbit dependent perturbation will deteriorate the formation design over time. In the VLEO range, any difference in the aerodynamic forces experienced by two satellites flying in formation, denoted as differential aerodynamic forces in the following, induces variations on their relative orbital elements. Whereas this is generally considered as an unwanted perturbing effect, it can be exploited for formation control purposes.\\

Whilst by the fundamental definition of the aerodynamic drag force the control authority of differential drag is mainly restricted to the in-plane relative motion control, differences in the respective lift forces can be used to alter the out-of-plane relative motion. In this case, the differences can be generated via differences in magnitude and/or the effective lift direction $ \hat{\boldsymbol{u}}_{L} $, which can be adjusted via the normal vectors n of the respective surfaces. An adjustment of the surface normal vectors can either be accomplished by rotating designated external panels, for example solar panels, or by changing the attitude of a non-symmetrical satellite. As aerodynamic lift (and therefore also differential lift) has so far mostly been neglected, commanding deviations in the pitch angles $ \theta_{C,D} $ of the two asymmetrically shaped satellites flying in formation are the method of choices to create differential drag. The thereby generated lift force, however, is also located within the orbital plane \cite{Walther.2020,Buhler.2021}.\\

\begin{figure}
	\centering
	\includegraphics[width=1\linewidth]{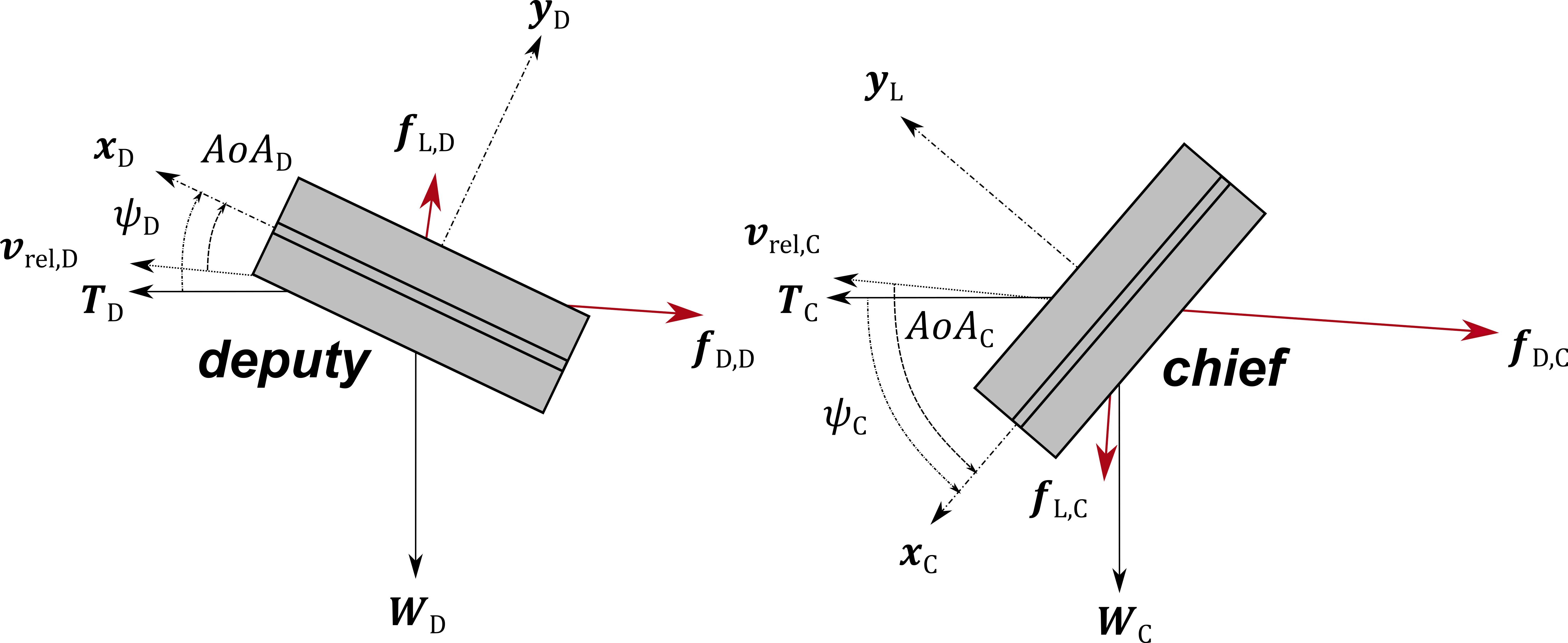}
	\caption{Visualization of a simultaneous differential lift and drag configuration via yaw angle deviations.}
	\label{fig:YawAngleDev}
\end{figure}

In the planning tool presented within this article, deviations in the yaw angles $ \psi_{C,D} $ of the respective satellites are applied to simultaneously create differential drag forces within the orbital plane and differential lift forces perpendicular to it (see also \cite{Hu.2021}). Thereby, all translational degrees of freedom of a satellite formation orbiting in VLEO can be controlled simultaneously without the need for any trusting device. A conceptual visualization of the proposed control approach, based on the coordinate frames introduced in subchapter \ref{ssec:ReferenceFrames}, is depicted in Fig. \ref{fig:YawAngleDev}.

\subsubsection{Free molecular flow / gas-surface interactions}
\label{sssec:FMF}
The residual atmosphere above 200 km is so rarefied that the mean free path of the gas molecules strongly exceeds the typical dimensions of a satellite \cite{Roberts.2017}. Thus, it cannot be considered a continuous fluid but a free molecular flow (FMF). In this regime, the so-called regime of extreme rarefaction \cite{S.A.Schaaf.1958}, the residual atmospheric gas needs to be considered particulate in nature and features negligible few collisions between constituent molecules. Thus, the incident flow can be assumed entirely undisturbed by the presence of the body \cite{S.A.Schaaf.1958}. As a consequence, the forces and torques occurring on a free body under FMF conditions are a result of the energy exchange taking place between the incident gas particles and the external surfaces. These gas–surface interactions (GSI) are dominated by the material chemistry with the predominant gas species in the VLEO range, atomic oxygen (AO), adsorbing to, and possibly eroding, the surface.\\

To best possibly estimate the aerodynamic coefficients $ C_{D} $ and $ C_{L} $ of satellites, models which adequately depict these GSIs are required. Within this article, Sentman’s GSI model \cite{Sentman.1961} is the model of choice as \textit{“[…] it is the de-facto standard to compute spacecraft aerodynamic coefficients at low altitude”} \cite{JosepVirgiliLlop.2014}. The model assumes that all incident particles colliding with the surface are adsorbed and later reemitted diffusely with a Maxwell–Boltzmann distribution of velocities and at a partial thermal equilibrium with the surface. How close the kinetic energy of the incoming molecules are adjusted to the thermal energy of the surface is expressed in the thermal accommodation coefficient $ \alpha_{T} $, which can be included in Sentman’s Equations as an input parameter via a method proposed by Moe et al. \cite{Moe.2005,Moe.2004} and later on slightly corrected by Koppenwallner \cite{Koppenwallner.2009}. To calculate $ \alpha_{T} $ with respect to the current environmental conditions, the ‘Semiempirical model for satellite energy accommodation coefficients’ (SESAM) developed by Pillinski et al. \cite{Pilinski.2010,Pilinski.2011} is applied. Following the physical explanation for accommodation, namely atomic oxygen adsorption, the model uses the partial pressure of AO as an input. Explicit equations for Sentman’s model are only available for a limited set of simple shapes, so that the drag and lift coefficients of satellite geometries of arbitrary (convex) shapes have to be approximated using the so-called \textit{panel method}. Here, the force coefficients of small surface elements are calculated in a finite element surface mesh and combined to form the drag coefficient of the entire satellite \cite{Pilinski.2011}. Within the planning tool, the panel method in combination with Sentman’s GSI model is the method of choice for calculating a reference for the aerodynamic coefficients of the satellites.

\subsection{Reference frames and Euler angle definition}
\label{ssec:ReferenceFrames}
This subchapter provides an overview of the reference frames and the definition of the yaw angle $ \psi $ used to develop the maneuver strategy.

\subsubsection{Earth-centered frames}
The Geocentric Equatorial coordinate system of Mean Equator and Equinox of J2000 (EME2000) (also referred to as Earth Centered Inertial (ECI) coordinate system) with its unit vectors $ \hat{\boldsymbol{I}} $, $ \hat{\boldsymbol{J}} $ and $ \hat{\boldsymbol{K}} $ is the most fundamental coordinate system to discuss Keplerian orbits. The system has its origin at the Earth’s center and the fundamental plane is its equator. The $ \hat{\boldsymbol{I}} $-axis points towards the vernal equinox, the $ \hat{\boldsymbol{J}} $-axis is tilted 90° to the east in the equatorial plane and the $ \hat{\boldsymbol{K}} $-axis extends through the North Pole. Orbital elements are expressed in the true-of-date (ToD) – frame, which is analogous to the ECI frame but accounts for the effect of precession and nutation (the $ \hat{\boldsymbol{I}}_{ToD} $ and $ \hat{\boldsymbol{K}}_{ToD} $-axes point towards the true equinox and north pole at the current epoch).

\subsubsection{Local vertical, local horizontal}
The coordinate system generally used to describe the relative motion of two spacecraft is the local vertical, local horizontal (LVLH) frame (also radial-transversal-normal or Hill frame). The unit vector triad is defined as $ \hat{\boldsymbol{x}} $, $ \hat{\boldsymbol{y}} $ and $ \hat{\boldsymbol{z}} $. The $ \hat{\boldsymbol{x}} $-axis points from the Earth’s center along the radius vector towards the satellite as it moves through the orbit. The $ \hat{\boldsymbol{y}} $-axis is perpendicular to the radius vector and points in the direction of (not necessarily parallel to) the velocity vector. The $ \hat{\boldsymbol{z}} $-axis is perpendicular to the orbital plane and aligned with the orbit angular momentum vector $ \boldsymbol{h} $. The mathematical expression is:
\begin{equation} \label{eq:LVLH}
		\hat{\boldsymbol{x}} = \frac{\boldsymbol{r}_{sat}}{|\boldsymbol{r}_{sat}|}, \ 
		\hat{\boldsymbol{z}} = \frac{\boldsymbol{r}_{sat} \times \boldsymbol{v}_{sat}}{|\boldsymbol{r}_{sat} \times \boldsymbol{v}_{sat}|}, \ 
		\hat{\boldsymbol{y}} = \hat{\boldsymbol{z}} \times \hat{\boldsymbol{x}}\\
\end{equation}

\subsubsection{Frenet system}
The Frenet system is defined such that the $ \hat{\boldsymbol{T}} $-axis points along the inertial velocity vector (tangential to the orbit), $ \hat{\boldsymbol{N}} $ lies in the orbital plane normal to the velocity vector and the $ \hat{\boldsymbol{W}} $-axis is normal to the orbital plane (parallel to $ \hat{\boldsymbol{z}} $) \cite{Vallado.op.2013}. This coordinate system is frequently employed to analyze the effect of aerodynamic drag as, neglecting perturbing effects (assuming $ \boldsymbol{v}_{rel} = \boldsymbol{v}_{sat}$), drag always acts antiparallel to the velocity vector and therefore in the -$ \hat{\boldsymbol{T}} $ direction. For the Frenet system, the mathematical expression of the unit vector triad is:
\begin{equation} \label{eq:Frenet}
		\hat{\boldsymbol{T}} = \frac{\boldsymbol{v}_{sat}}{|\boldsymbol{v}_{sat}|}, \
		\hat{\boldsymbol{W}} = \frac{\boldsymbol{r}_{sat} \times \boldsymbol{v}_{sat}}{|\boldsymbol{r}_{sat} \times \boldsymbol{v}_{sat}|}, \
		\hat{\boldsymbol{N}} = \hat{\boldsymbol{T}} \times \hat{\boldsymbol{W}}\\
\end{equation}

\subsubsection{Body fixed}
The body fixed (BF) frame is centered at the center of mass of the satellite and the $ \hat{\boldsymbol{x}}_{BF} $, $ \hat{\boldsymbol{y}}_{BF} $, and $ \hat{\boldsymbol{z}}_{BF} $-axes are aligned with its principal axes forming a right-handed frame. Throughout this article, the nominal attitude of the satellites (i.e. if no relative motion control is applied) is defined in a way that the BF – frame tracks the Frenet - frame so that the satellite’s $ \hat{\boldsymbol{x}}_{BF} $ is aligned with the $ \hat{\boldsymbol{T}} $-axis, the $ \hat{\boldsymbol{z}}_{BF} $-axis is pointing towards $ -\hat{\boldsymbol{N}} $ and $ \hat{\boldsymbol{y}}_{BF} $ completes the right handed frame (points along $- \hat{\boldsymbol{W}} $ ).

\subsubsection{Yaw angle $ \psi $ definition}
For control purposes, a desired yaw angle $ \psi $ can be commanded to each satellite. Within this article, the Euler angles ($ \phi $ / $ \theta $ / $ \psi $) are defined with respect to the Frenet–frame. The yaw angle $ \psi $ is defined as the angle between the $ \hat{\boldsymbol{x}}_{BF} $–axis of the body fixed frame and the $ \hat{\boldsymbol{T}}$–axis of the Frenet frame due to a rotation around $ \hat{\boldsymbol{z}}_{BF} $. Consequently, for a yaw angle of $ \psi = 0^{\circ} $ the satellite’s $ \hat{\boldsymbol{x}}_{BF} $–axis is aligned with its inertial velocity vector $\boldsymbol{v}_{sat}$.\\ 

\textbf{Note}: due to the co-rotation of the atmosphere with the Earth and the presence of thermospheric winds, the velocity relative to the local atmosphere deviates from the satellite velocity ($ \boldsymbol{v}_{rel} \neq \boldsymbol{v}_{sat} $) so that there is a discrepancy between the pitch angle $ \psi $ and the angle-of-attack (AoA) of the satellite, which is defined as the angle between the $ \hat{\boldsymbol{x}}_{BF} $–axis and $\boldsymbol{v}_{rel}$. A visualization of the difference between the angle-of-attack and the yaw angle definition can be found in Fig. \ref{fig:YawAngleDev}.

\subsection{Nearly-nonsingular orbital elements}
\label{ssec:NNOE}
For a simple geometrical representation, Keplerian elements $ \boldsymbol{\mathcal{E}} = (a, \ e, \ i, \ \Omega, \ \omega, \ \theta)^{T} $ are commonly used to describe the state of a satellite. Here, $ a $ is the semi-major axis, $ e  $ is the eccentricity, $ i $ is the inclination, $ \Omega $ is the right ascension of the ascending node, $ \omega $ is the argument of perigee and $\theta$ is the true anomaly, which is replaceable by the mean anomaly  $ M $ \cite{Battin.2000}. As the Keplerian elements are singular for circular and equatorial orbits, nearly-nonsingular (ns) mean orbital elements are employed within this article. Using the definition of the Keplerian elements $ \boldsymbol{\mathcal{E}} $, these are formally defined as:
\begin{equation}\label{eq:NearNonsingularElements}
	\begin{split}
		\boldsymbol{\mathcal{E}}_{ns} = \biggl(a,\ \lambda,\ i,\ q_{1} = e \cos(\omega), \ q_{2} = e \sin(\omega),\ \Omega     \biggr)^{T}
	\end{split}
\end{equation}
where $ \lambda = M + \omega $ is the mean argument of latitude, frequently also replaced by the true argument of latitde $ u = \omega + \theta $. Notably, this set of orbital elements is still singular for equatorial orbits.\\ 

Within this article, the motion of the deputy with respect to the chief is described by the following set of mean differential orbital elements \cite{Roscoe.2015}:
\begin{equation} \label{eq:dNNS}
 \delta \bar{\boldsymbol{\mathcal{E}}}_{ns}  = \bar{\boldsymbol{\mathcal{E}}}_{ns,D} - \bar{\boldsymbol{\mathcal{E}}}_{ns,C}  
\end{equation}
By specifying the relative orbit geometry in mean element space, the true relative spacecraft motion closely follows the prescribed relative orbit geometry \cite{Schaub.2000b}. Note that the relative obit description from Eq. \ref{eq:dNNS} does not make any assumptions on how large the relative orbit is compared to the chief orbit radius, nor does it require the chief orbit to be circular \cite{Schaub.2018}. 

\subsection{Characterization of satellite relative motion}
\label{ssec:CW}
In a general unperturbed elliptic orbit, the motion of the deputy relative to the chief in the LVLH frame is governed by the Tschauner-Hempel (TH) \cite{Tschauner.1965} or Lawden's \cite{Lawden.1963} equations. A parametrization of the general solution to these equations in terms of nearly-nonsingular elements has been derived by Sengupta and Vadali and is recited for reference \cite{Sengupta.2007,Roscoe.2015}:
\begin{equation} \label{eq:TH1}
	\begin{split}
		x(u) = \rho_{1}\sin(u+\tilde{\alpha}_{0}) \\
		- \frac{2v_{d}}{3n\eta^2} \left[\frac{r}{p} - \frac{3\left(q_{1} \sin(u) - q_{2} \cos(u) \right) }{2\eta^3} K(u)\right] 
	\end{split}
\end{equation}
\begin{equation} \label{eq:TH2}
	\begin{split}
		y(u) = \frac{\rho_{1}r}{p}\left(2+q_{1} \cos(u) + q_{2} \sin(u) \right) \cos(u + \tilde{\alpha}_{0})\\
		+ \frac{\rho_{2}r}{p} + \frac{v_{d}p}{rn\eta^5}K(u)
	\end{split}
\end{equation}
\begin{equation} \label{eq:TH3}
	z(u) = \frac{\rho_{3}r}{p} \sin(u + \tilde{\beta}_{0})
\end{equation}
where $ K( u ) $ is an implicit function of the true argument of latitde $ u $:
\begin{equation} \label{eq:TH4}
	K( u ) = \lambda - \lambda_{0} = n(t-t_{0})
\end{equation}
In Eq. \ref{eq:TH1}-\ref{eq:TH4}, $ p = a\eta^{2} $ is the semiparameter, $ \eta = \sqrt{1-e^{2}} $ and $ n = \sqrt{\frac{\mu_{e}}{a^{3}}}$ the mean motion. $\boldsymbol{\rho} =(x,y,z)^{T} $ are the components of the relative position vector $ \boldsymbol{\rho} $  of the deputy with respect to the chief in the LVLH frame and $ \rho_{1} $, $ \rho_{2} $, $ \rho_{3} $, $ v_{d} $, $ \tilde{\alpha}_{0} $, and $ \tilde{\beta}_{0} $ are the parameters that define the relative trajectory. $ \rho_{1} $ and $ \rho_{3} $ relate to the amplitude of the in-plane and out-of-plane motion, $ \alpha_{0} $ and $ \beta_{0} $ are the initial phase angles, $ \rho_{2} $ determines how far offset the motion is in the along-track direction, and $ v_{d} $ is the along-track drift rate. The advantage of this parameterization is their uniform validity for all eccentricities, and the fact that the effects of changing one or more of these parameters is intuitively clear \cite{Sengupta.2007}. \\

\begin{figure}
	\centering
	\includegraphics[width=\linewidth]{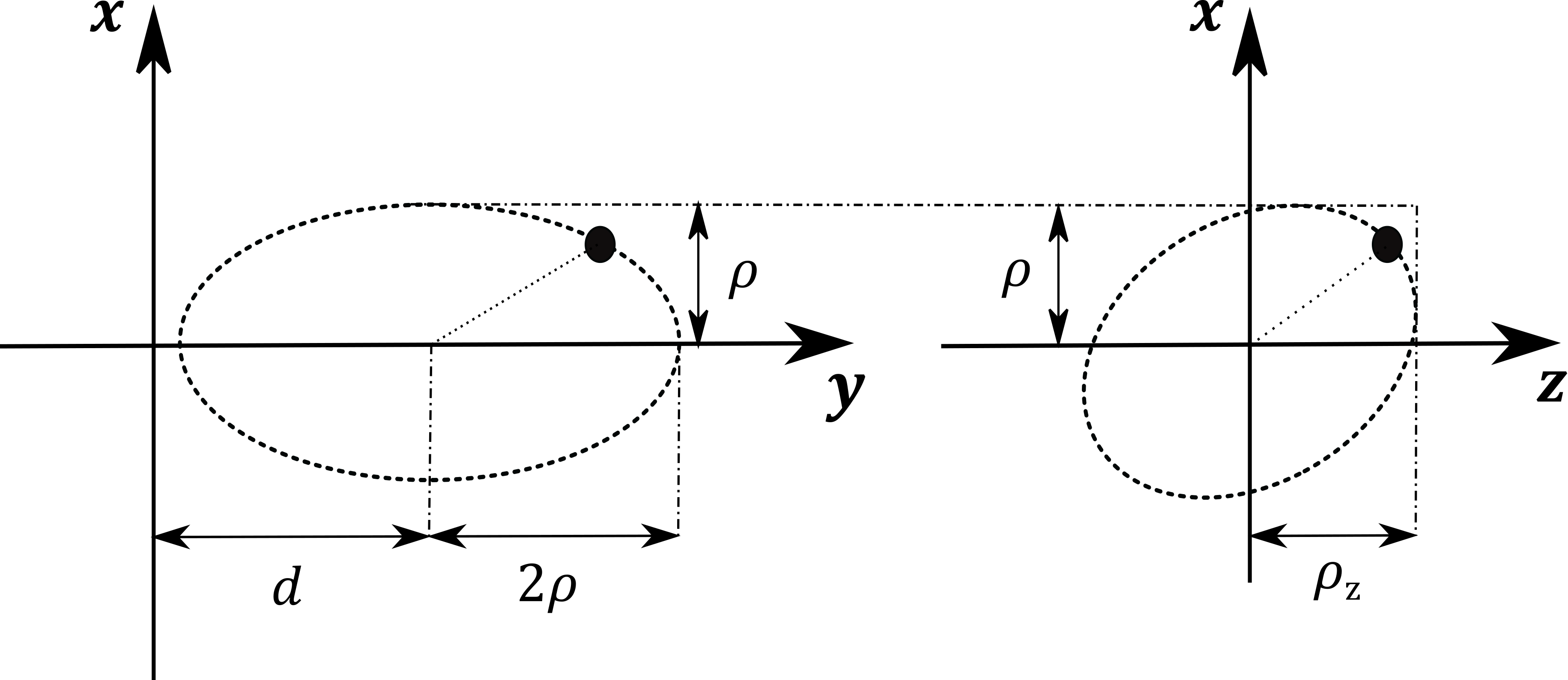}
	\caption{Projections of the relative motion solution of the CW equations in the along-track/radial (left) and cross-track/radial (right) directions for an arbitrary in- and out-of-plane oscillating formation design with zero radial offset $ da = 0 $.}
	\label{fig:CW}
\end{figure}

For the circular chief orbit special case, Eqs. \ref{eq:TH1}-\ref{eq:TH3} simplify to the well-known Hill-Clohessy-Wiltshire (HCW) \cite{ClohessyWH.1960,Hill.1878} equations:
\begin{equation} \label{eq:CW1}
	x(t) = \rho \sin(nt+\alpha_{0}) - \frac{2v_{d}}{3n}
\end{equation}
\begin{equation} \label{eq:CW2}
	y(t) = 2 \rho \cos(nt+\alpha_{0}) + d + v_{d}(t-t_{0})
\end{equation}
\begin{equation} \label{eq:CW3}
	z(t) = \rho_{z} \sin(nt+\beta_{0}) 
\end{equation}
where the following relations with respect to the parameters of the TH equations hold: $ \rho_{1} = \rho $, $ \rho_{2} = d $, $\rho_{3} = \rho_{z}$, $\tilde{\alpha}_{0} = \alpha_{0} $ and $\tilde{\beta}_{0} = \beta_{0} $. Consequently, in the circular chief orbit case, $ \rho $ and $ \rho_{z} $ correspond precisely to the amplitude of the in-plane and out-of-plane motion, the in-plane motion is the superposition of a 2-1 ellipse and a linear drift in the along-track direction at a rate of $ v_{d} $, and $ d $ is the along-track offset of the initial in-plane ellipse. Any along-track drift also induces a small constant radial offset. The parameters are visualized for an arbitrary in- and out-of-plane oscillating formation design with zero radial offset in Fig. \ref{fig:CW}.

Passively safe trajectories can be accomplished by applying a well chosen superposition of the in-plane and the out-of-plane harmonic motion with initial phase angles selected so that the trajectory never crosses the along-track axis. In this way, an un-intentional drift of the deputy towards the chief in the along-track direction would not lead to a collision since the deputy never actually crosses the along-track axis. Maximum "safety" is achieved when the phase angles are chosen such that $ \beta_{0}= \alpha_{0} + \frac{\pi}{2} $ since this maximizes the distance between the deputy and the alongtrack axis when either $ x = 0 $ or $ z = 0 $, which is ensured in any case hereinafter \cite{Roscoe.2015}. 

\subsection{Mapping between LVLH states and mean differential orbital elements}
\label{ssec:Mapping}
To use the geometrical insights provided by the TH (or HCW) equations for the definition of the initial and final relative states of a maneuver under investigation, relationships to convert the respective parameters into nearly-nonsingular element differences are required. These are provided by Sengupta and Vadali \cite{Sengupta.2007} and summarized here for reference\footnote{\textbf{Note:} With the mapping shown above (Eq. \ref{eq:Map1}-\ref{eq:Map6}), the desired parameters are mapped into mean orbital elements $ \delta \bar{\boldsymbol{\mathcal{E}}}_{ns} $ at $ n t_{0} = \lambda_{0} = 0 $.} \cite{Roscoe.2015}:
\begin{equation} \label{eq:Map1}
\delta a = -\dfrac{2\eta v_{d}}{3n}
\end{equation}
\begin{equation} \label{eq:Map2}
\delta \lambda_{0} = \frac{\rho_{2}}{p}-\delta \Omega \cos(i) - \frac{1+\eta+\eta^2}{1+\eta} \frac{\rho_{1}}{p} (q_{1} \cos(\tilde{\alpha}_{0})- q_{2} \sin(\tilde{\alpha}_{0}))
\end{equation}
\begin{equation} \label{eq:Map3}
\delta i = \frac{\rho_{3}}{p} \cos(\tilde{\beta}_{0})
\end{equation}
\begin{equation} \label{eq:Map4}
	\delta q_{1} = -(1-q_{1}^{2})\frac{\rho_{1}}{p}\sin(\tilde{\alpha}_{0}) + q_{1}q_{2}\frac{\rho_{1}}{p}\cos(\tilde{\alpha}_{0}) - q_{2}\left(\frac{\rho_{2}}{p} - \delta \Omega \cos(i)\right) 
\end{equation}
\begin{equation} \label{eq:Map5}
		\delta q_{2} = -(1-q_{2}^{2})\frac{\rho_{1}}{p}\cos(\tilde{\alpha}_{0}) + q_{1}q_{2}\frac{\rho_{1}}{p}\sin(\tilde{\alpha}_{0}) + q_{1}\left(\frac{\rho_{2}}{p} - \delta \Omega \cos(i)\right) 
\end{equation}
\begin{equation} \label{eq:Map6}
	\delta \Omega = -\frac{\rho_{3}}{p}\frac{\sin(\tilde{\beta}_{0})}{\sin(i)}
\end{equation}
The necessary conditions for a bounded, centered relative motion of a deputy with respect to the chief spacecraft are given by:
\begin{equation} \label{eq:BCRM1}
\delta a = 0 
\end{equation}
\begin{equation} \label{eq:BCRM2}
\delta \lambda_{0} = - \delta \Omega \cos(i)
\end{equation}
In this case, the relative orbit of the deputy with respect to chief spacecraft is an ellipse of semi-major axis $ 2\rho $ in along-track direction and semi-minor axis $ \rho $ in radial direction (see Fig. \ref{fig:CW})\footnote{Again, for quasi-circular formations the following  relations hold: $ \rho_{1} = \rho $, $ \rho_{2} = d $, $\rho_{3} = \rho_{z}$, $\tilde{\alpha}_{0} = \alpha_{0} $ and $\tilde{\beta}_{0} = \beta_{0} $.}. While $ \rho $ measures the size of the relative trajectory, the angle $ \alpha_{0} $ defines the relative pericenter. Whenever the sum of the argument of latitude $ u $ and $ \alpha_{0} $ equals $ \frac{\pi}{2} $, the deputy is located right above the center of the 2-1 ellipse. As soon as $ u + \alpha_{0} = \pi $, the deputy is at the maximum distance behind the chief satellite with zero radial distance. At $ u + \alpha_{0} = \frac{3\pi}{2} $, the deputy is below the chief and at $ u + \alpha_{0} = 2\pi$ right in front of the chief. The out-of-plane relative motion is further described by a harmonic oscillation of amplitude $ \rho_{z} $ and phase angle $ u + \beta_{0} $. \\

Notably, the methodology enables to define common formation designs, such as circular (CF) or projected circular formation (PCF) (following the developments by Sabol et al. \cite{Sabol.2001}), in terms of nearly-nonsingular orbital elements. Moreover, statements regarding the relative orbit geometry can be derived from the orbital element differences. As an example,  the magnitude of the out-of-plane relative motion is a direct result of differences in the inclination and in the ascending nodes. Whereas differences in the inclination angle  $ \delta i $ specify how much out-of-plane motion the relative orbit will have as the satellite crosses the northern- or southernmost regions, ascending node differences $ \delta \Omega $, however, indicates the out-of-plane motion as the satellite crosses the equatorial plane (at the ascending node) \cite{Schaub.2018}. 

\section{Optimal maneuver planning tool}
\label{sec:PlanningTool}
The optimal control approach proposed in \cite{DellElce.2015,DellElce.2015b}, which aimed at the planning and execution of an in-plane rendezvous maneuver between an active deputy as well as a non-cooperative target satellite via differential drag, consisted of three phases: in a first phase, the ballistic coefficient $ \beta $ of the active satellite was fitted in-orbit for different pitch angles $ \theta $ via a least mean squared error approach. In a second step, the maneuver was planned using a numerical optimizer. In a third and final phase, the maneuver was executed while an on-line compensator accounted for uncertainties and un-modeled dynamics and thereby ensured a proper tracking of the scheduled trajectory. Whereas promising control options, i.e. adaptive control techniques, able to deal to cope with the dynamic variations, uncertainty and noise with manageable demands on the available computing resources exist and an application of which to the underlying problem is of highest interest to the community, the desire of this article is not to propose a new closed-loop control approach but a flexible, powerful and high-fidelity planning tool for maneuver sequences targeting optimal behavior. Therefore, at this point we would like to separate these two research efforts and focus on the second phase only, namely the maneuver planning process, within this article. The rationale behind this decision is that it is our firm believe that a flexible and powerful planning tool is required to enable parameter studies to explore and outline the design space of possible maneuver variants and thus move the methodology from a promising theoretical concept to an actual viable control option for a wide range of satellite missions. A focusing on the planning process, however, allows further to increase the model fidelity of the planning tool by incorporating additional effects, such as the co-rotation of the atmosphere with the Earth.\\

In this subchapter, the complete theory implemented in the maneuver planner is presented and discussed in detail.
\subsection{Pre-processing}
\label{ssec:Preproc}
The computational burden of the planning process can be significantly decreased with only a minor loss in fidelity if some pre-processing prior to the maneuver planning process is performed. In this subchapter, the performed steps are discussed and justified.

\subsubsection{Fitting of the coefficients of the analytic density model}
\label{ssec:DensityModelFit}
The arguably most significant limitation of the simplified maneuver trajectories \cite{Walther.2020,Buhler.2021} in terms of the achievable accuracy is their underlying constant density assumption as, in reality, the density of the upper atmosphere is known to be subjected to variations due to a complex interaction between the nature of the atmosphere’s molecular structure, the incident solar flux as well as geomagnetic (auroral) interactions \cite{Vallado.op.2013}. To account for those dynamic variations within numerical orbit propagations, empirical density models are commonly the method of choice. An incorporation of such a model in the planning process, however, would significantly slow down the calculation. For this type of study, that it does not try to use the density results to plan or study any operational mission, a sufficiently adequate representation of the density along the maneuver requiring only a fraction of the computational burden can be obtained via the following analytic density model \cite{DellElce.2015,DellElce.2015b}:
\begin{equation} \label{eq:DenstyModel}
	\rho = A \left(1+B \cos\left( u-C\right) \right) \exp \left(\frac{r-R_{e} \ \sqrt{1-e_{e}^{2} \sin(i)^{2} \sin(u)^{2}}}{D} \right) 
\end{equation}
for which the four coefficients ($ A $,$ B $,$ C $,$ D $) are orbit and epoch dependent and fitted to the NRLMSISE-00 environmental model \cite{Picone.2002} prior to initiating the planning process. The fitting is performed in a sense that the mean squared error (MSE):
\begin{equation}\label{eq:MSE}
	MSE = \frac{1}{n} \sum_{i=1}^n \left(\rho_{i,AM} - \rho_{i,ref} \right)^{2} 
\end{equation}
between the values predicted by the analytic model (subscript \textit{AM}) (Eq. \ref{eq:DenstyModel}) and the NRLMSISE-00 environmental model \cite{Picone.2002} (subscript \textit{ref}) is minimized. While fitting the model coefficients accounts for influences such as the epoch and the solar and geomagnetic activities, the model itself is able to take the most relevant characteristics of the upper atmosphere, namely the exponential vertical structure, the day/night bulge, and the Earth’s oblateness, into account. In Eq. \ref{eq:DenstyModel}, $ u $ is the true argument of latitude, $ r $ is the distance from the Earth's center to the spacecraft, $ i $ is the inclination, $R_{e} $ is the Earth's mean equatorial radius and $ e_{e} $ the Earth's eccentricity.

\subsubsection{Fitting of the aerodynamic properties of the spacecraft}
\label{ssec:AeroPropFit}
Whereas, compared to more accurate numerical methods, such as the Direct Simulation Monte Carlo (DSMC) or the Test-Particle Monte Carlo (TPMC) method, the computational burden of the panel method is significantly reduced, an evaluation of the respective set of equations of the GSI model is required for each plate of the surface mesh and each time step. In addition, this requires a knowledge of the current environmental conditions which again necessitates an inclusion of a computational expensive environmental model. To adapt the aerodynamic calculations to the satellites under investigations while keeping the computational burden limited, in a second pre-processing step the ballistic coefficient $ \beta $ and its respective counterpart for lift $ \beta_{L} $ are fitted for both satellites as a function of their angle-of-attack AoA to the respective reference data which is produced via the methodology described in subchapter \ref{ssec:AeroPropFit}. All required environmental input parameters, i.e. the thermospheric temperature $ T $, the mean molecular mass $ \bar{M} $ and the particle number density of atomic oxygen $ n_{O} $, are included as orbit averaged values for the particular orbit under investigation and calculated via the NRLMSISE-00 environmental model \cite{Picone.2002}.\\

The result of this pre-processing step are the coefficients $ \beta = f(AoA) $ and $ \beta_{L} = f(AoA) $ as functions of the AoA for both satellites, which are used to estimate the respective values within the optimizer.

\subsection{Maneuver planner}
\label{ssec:ManPlan}
The optimal control aims to drive the dynamical system, $ \boldsymbol{\dot{x}}=f(\boldsymbol{x},\boldsymbol{u},t) $ in which $ \boldsymbol{x} $ are the states and $ \boldsymbol{u} $ the control variables, from a initial state $ \boldsymbol{x}(0) = \boldsymbol{x}_{0} $ to a desired final state $ \boldsymbol{x}(t_{f}) = \boldsymbol{x}_{f} $ with $ t_{f} $ being the maneuvering time. The trajectory is optimized in a sense that it minimizes a cost functional $ \mathcal{J}(\boldsymbol{x},\boldsymbol{u},t_{f})$:
\begin{equation}
	\mathcal{J}(\boldsymbol{x},\boldsymbol{u},t_{f}) = \mathcal{M}(t_{f}) + \int\limits_{0}^{t_{f}} \mathcal{L}(\boldsymbol{x},\boldsymbol{u},t) \ dt
\end{equation} 
while satisfying the inequality constraints $ g(\boldsymbol{x},\boldsymbol{u},t) \leq 0  $. In summary, the set of constraints yields the following \textit{Bolza} problem \cite{DellElce.2015b}:
\begin{equation}\label{eq:Bolza}
	\begin{split}
		\lbrack \boldsymbol{x}^{*}, \boldsymbol{u}^{*} \rbrack = \mathrm{arg} \ \Bigl[ \underset{\boldsymbol{x(t)},\boldsymbol{u(t)},t \in \lbrack 0, t_{f} \rbrack}{\mathrm{min}} \ \mathcal{J}(\boldsymbol{x},\boldsymbol{u},t_{f}) \Bigr] \ \textit{s.t.}\\
		\boldsymbol{\dot{x}} = f(\boldsymbol{x},\boldsymbol{u},t) \ \ \ \ \forall t \in \lbrack 0, t_{f} \rbrack\\
    	g(\boldsymbol{x},\boldsymbol{u},t) \leq 0   \ \ \ \ \forall t \in \lbrack 0, t_{f} \rbrack\\
		\boldsymbol{x}(t_{0})= \boldsymbol{x}_{0} \\
		\boldsymbol{x}(t_{f}) = \boldsymbol{x}_{f} \\
	\end{split}
\end{equation}
In the following, the different quantities for the problem studied hereinafter are introduced.

\subsubsection{Dynamical system}
\label{sssec:DynSys}
The state vector $ \boldsymbol{x} $ considered in the control plant is:
\def\x{
	\begin{bmatrix}
		\delta \bar{\boldsymbol{\mathcal{E}}}_{ns}\\
	        	\bar{\boldsymbol{\mathcal{E}}}_{ns,C}    \\
		\psi_{C}\\
		\dot{\psi}_{C}\\
		\psi_{D}\\
		\dot{\psi}_{D}
\end{bmatrix}}
\begin{equation}
	\boldsymbol{x} =\x
\end{equation}
which consists of the mean differential nearly-nonsingular orbital elements $ \delta \bar{\boldsymbol{\mathcal{E}}}_{ns} $, the mean nonsingular elements of the chief $ \bar{\boldsymbol{\mathcal{E}}}_{ns,C} $ as well as the yaw angles $ \psi_{C,D} $ and the rotational velocity $ \dot{\psi}_{C,D} $ of both satellites.\\

With the mean nearly-singular elements of the chief $ \bar{\boldsymbol{\mathcal{E}}}_{ns,C} $ being included in the state vector $ \boldsymbol{x} $, the respective counterparts of the deputy $ \bar{\boldsymbol{\mathcal{E}}}_{ns,D} $ can simply be calculated via:
\begin{equation}
	\bar{\boldsymbol{\mathcal{E}}}_{ns,D}  =  \bar{\boldsymbol{\mathcal{E}}}_{ns,C}  + \delta \bar{\boldsymbol{\mathcal{E}}}_{ns}
\end{equation}
so that the full absolute and relative states of the two spacecraft are known at all times. From the mean non-singular elements, mean Keplerian elements $ \bar{\boldsymbol{\mathcal{E}}}  $ as well as in a second conversion step their osculating counterparts $ \boldsymbol{\mathcal{E}} $ can be obtained. In a last conversion, these can be mapped into the osculating inertial position $ \boldsymbol{r}_{sat} $ and $ \boldsymbol{v}_{sat} $ in the ToD-frame, which are used to calculate the aerodynamic forces acting on the satellites.\\

The relative motion dynamics is modeled using the linearized equations presented by Roscoe et al. \cite{Roscoe.2015}:
\begin{equation}
	\delta \dot{\bar{\boldsymbol{\mathcal{E}}}}_{ns}  =  \left[\boldsymbol{A}(\bar{\boldsymbol{\mathcal{E}}}_{ns,C}) \right] \delta \bar{\boldsymbol{\mathcal{E}}}_{ns}  + \left[\boldsymbol{B}(\bar{\boldsymbol{\mathcal{E}}}_{ns,C}) \right] \boldsymbol{u}
\end{equation}
where $ \left[\boldsymbol{B}(\bar{\boldsymbol{\mathcal{E}}}_{ns,C}) \right] $ are Gauss's Variational Equations (GVE) and $ \left[\boldsymbol{A}(\bar{\boldsymbol{\mathcal{E}}}_{ns,C}) \right] $ is the Jacobian of Lagrange's Planetary Equations (LPE) evaluated on the reference orbit. For reference, the equations for  $ \left[\boldsymbol{B}(\bar{\boldsymbol{\mathcal{E}}}_{ns,C}) \right] $  and of  $ \left[\boldsymbol{A}(\bar{\boldsymbol{\mathcal{E}}}_{ns,C}) \right] $ are recited in \ref{GVE} and \ref{DLPE}, respectively. \\

For the methodology studied within this article, the control vector $ \boldsymbol{u} $ of choice is the arithmetic difference of the specific aerodynamic forces acting on the satellites $ \delta \boldsymbol{f}_{aero} $ expressed in the LVLH - frame of the chief:
\begin{equation}
	\delta \boldsymbol{f}_{aero} = \boldsymbol{f}_{aero,D} - \boldsymbol{f}_{aero,C} = \left(f_{aero,x},\ f_{aero,y},\ f_{aero,z} \right)^{T}
\end{equation}
Since the gradient of the contract transformation is close to the identity matrix \cite{Schaub.2018,Schaub.2000b}, the instantaneous specific aerodynamic forces $ \boldsymbol{f}_{aero,D} $ and $ \boldsymbol{f}_{aero,D} $ are used to calculate this difference \cite{DellElce.2015b,Schaub.2000b}. These are calculated for each spacecraft via Eq. \ref{eq:Drag}, \ref{eq:Lift} and \ref{eq:Aero}, respectively. In Eq. \ref{eq:Drag} and \ref{eq:Lift}, the local atmospheric density is calculated using the analytic density model presented in subchapter \ref{ssec:DensityModelFit} and the relative velocity vector $ \boldsymbol{v}_{rel} $ is calculated via Eq. \ref{eq:RelVel}, in which wind effects are neglected ($ \boldsymbol{v}_{wind} = \boldsymbol{0} $). The sensitivity of the satellites towards the aerodynamic forces are incorporated in the equations via the functions for $ \beta = f(AoA) $ and $ \beta_{L} = f(AoA) $ discussed in subchapter \ref{ssec:AeroPropFit}. As by the definition of the yaw angle (see subchapter 2.2.5) the body fixed frame is rotated with respect to the Frenet – frame by an angle of $ \psi $ around the $ \hat{\boldsymbol{z}}_{BF} $–axis (which points along $ -\hat{\boldsymbol{N}} $), the orientation of the satellite body with respect to the flow, i.e. the angle-of-attack (the angle between the $ \hat{\boldsymbol{x}}_{BF} $–axis of the satellite and the velocity vector relative to the local flow $ \boldsymbol{v}_{rel} $), can be determined at all times. \\

The absolute motion dynamics of the chief, expressed via mean nearly-nonsingular orbital elements $ \bar{\boldsymbol{\mathcal{E}}}_{ns,C} $, is calculated using the Lagrange's Planetary Equations (LPE) and a modified form (to use the nearly-nonsingular elements) of Gauss's Variational Equations (GVE). The LPE provide a set of equations relating the effect of a control acceleration vector $ \boldsymbol{u} $ to the osculating orbit element’s time derivatives. Since GVE give the effect of accelerations on the osculating elements, the osculating-mean transformation must then be applied to determine changes in the mean elements. However, as the sensitivities of mean element changes with respect to osculating element changes are of at most $ \mathcal{O}(J_{2}) $, the following approximation is used \cite{Schaub.2000b,Schaub.2001}:
\begin{equation}
	\dot{\bar{\boldsymbol{\mathcal{E}}}}_{ns,C} \approx \boldsymbol{f}(\bar{\boldsymbol{\mathcal{E}}}_{ns,C}) + \left[\boldsymbol{B}(\bar{\boldsymbol{\mathcal{E}}}_{ns,C}) \right] \boldsymbol{f}_{aero,C}
\end{equation}
where the unforced dynamics $ \boldsymbol{f}(\bar{\boldsymbol{\mathcal{E}}}_{ns,C})  $ including $ J_{2} $ is given by:
\def\f{
	\begin{bmatrix}
		0\\
		n + \frac{3}{4} J_{2} \left( \frac{R_{e}}{p}\right)^{2} n \left[\eta (3 \cos(i)^{2}-1)+(5\cos(i)^2 -1) \right]    \\
		0\\
		-\frac{3}{4} J_{2} \left( \frac{R_{e}}{p}\right)^{2} n (3 \cos(i)^2-1) q_{2}\\
		+\frac{3}{4} J_{2} \left( \frac{R_{e}}{p}\right)^{2} n (3 \cos(i)^2-1) q_{1}\\
		-\frac{3}{2} J_{2} \left( \frac{R_{e}}{p}\right)^{2} n \cos(i)
\end{bmatrix}}
\begin{equation}
	\boldsymbol{f}(\bar{\boldsymbol{\mathcal{E}}}_{ns,C})  =\f
\end{equation}

The rotational dynamics of both spacecraft around their $ \hat{\boldsymbol{z}}_{BF} $–axes are included in a simplified form neglecting any perturbing effects and cross-term couplings. These are included to ensure a smooth and realistic profile of the yaw angles $ \psi_{C,D}(t) $ which avoid bang-bang type control switches:\\
\begin{equation}
	\ddot{\psi}_{C,D} = -I_{sat,z}u_{C,D}
\end{equation}
Here, $ I_{sat,z} $ is the satellite’s moment of inertia around the $ \hat{\boldsymbol{z}}_{BF} $–axis and $ u_{C,D} $ the actual control variables in the optimizer, are the torques commanded to the reaction wheel of the respective satellite. Notably, due to the underlying simplifications included in the rotational dynamics, not the resulting torque but the resulting yaw angle profiles $ \psi_{C}^{\star} $ and $ \psi_{D}^{\star} $ are considered the results of the planner. In a real maneuver application, it would be the task of the attitude control systems of the satellites to ensure a proper tracking of the profile. \\

\textbf{Result}: the final output of the maneuver planner are the reference states $ \boldsymbol{x}^{\star} $ as well as the resulting yaw angle profiles $ \psi_{C}^{\star} $ and $ \psi_{D}^{\star} $ as a function of time.

\subsubsection{Cost function}
\label{sssec:Costs}
The calculated control profile is optimal in a sense that it minimizes a desired cost functional $ \mathcal{J}(\boldsymbol{x},\boldsymbol{u},t_{f})$. Since the goal is to minimize the loss in specific mechanical energy during the maneuver, the final mean semi major axis of the chief $ \bar{a}_{C,f} $ is maximized:
\begin{equation}
	\mathcal{J}(\boldsymbol{x},\boldsymbol{u},t_{f}) = -\bar{a}_{C,f}
\end{equation}

\subsubsection{Constraints}
\label{sssec:Constraints}
Constraints included are the admissible value range of the yaw angle of both satellites $ \psi_{C,D} $, their minimum and maximum angular velocity $ \dot{\psi}_{C,D}(t) $ as well as the absolute value of the maximum torque commanded to the reaction wheels $ T_{W,max} $: 
\begin{equation}\label{eq:Constraints}
	\psi_{C,D}  \in \lbrack \psi_{min}, \ \psi_{max}\rbrack  \ \  \forall t \in \lbrack 0, t_{f} \rbrack\\
\end{equation}
\begin{equation}\label{eq:Constraints2}
	\dot{\psi}_{C,D}  \in \lbrack \dot{\psi}_{min}, \ \dot{\psi}_{max}\rbrack \ \  \forall t \in \lbrack 0, t_{f} \rbrack\\
\end{equation}
\begin{equation}\label{eq:Constraints3}
	u_{C,D}  \in \lbrack -T_{w,max}, \ T_{w,max} \rbrack  \ \  \forall t \in \lbrack 0, t_{f} \rbrack\\
\end{equation}
for $ t_{f}  \in \lbrack 	t_{f,min}, \ t_{f,max} \rbrack  $. An exact value for the maneuver time $ t_{f} $ is not prescribed in order to broaden the solution space. However, limits are included to facilitate the solution finding process.

During the maneuver, no constraints on the absolute $ \dot{\bar{\boldsymbol{\mathcal{E}}}}_{ns,C} $ or relative states $ \delta \dot{\bar{\boldsymbol{\mathcal{E}}}}_{ns} $ are set. However, note that the initial absolute $ \dot{\bar{\boldsymbol{\mathcal{E}}}}_{ns,C,0} $ and relative states $ \delta \dot{\bar{\boldsymbol{\mathcal{E}}}}_{ns,0} $ as well as the desired relative formation geometry after the maneuver $ \delta \dot{\bar{\boldsymbol{\mathcal{E}}}}_{ns,f} $ are fixed. The final absolute state of the chief $ \dot{\bar{\boldsymbol{\mathcal{E}}}}_{ns,C,f} $, however, remain unconstrained. 

\subsubsection{Solution}
\label{sssec:Solution}
The solution to the problem is obtained by using collocation which is performed using the software GPOPS-II \cite{Patterson.2014}. GPOPS-II employs an hp-adaptive version of the Legendre-Gauss-Radau (LGR) orthogonal collocation method, a Gaussian quadrature implicit integration method where collocation is performed at the LGR points. The optimization is carried out means of the nonlinear programming solver IPOPT \cite{Wachter.2006}. To avoid numerical issues, all translational states besides $ \delta a $ and $ a_{C} $ are multiplied with the Earth's mean equatorial radius.

\section{Optimal maneuver trajectories}
\label{sec:Trajectories}

\subsection{Maneuver setup}
\label{ssec:ManSetup}
\begin{figure}
	\centering
	\includegraphics[width=\linewidth]{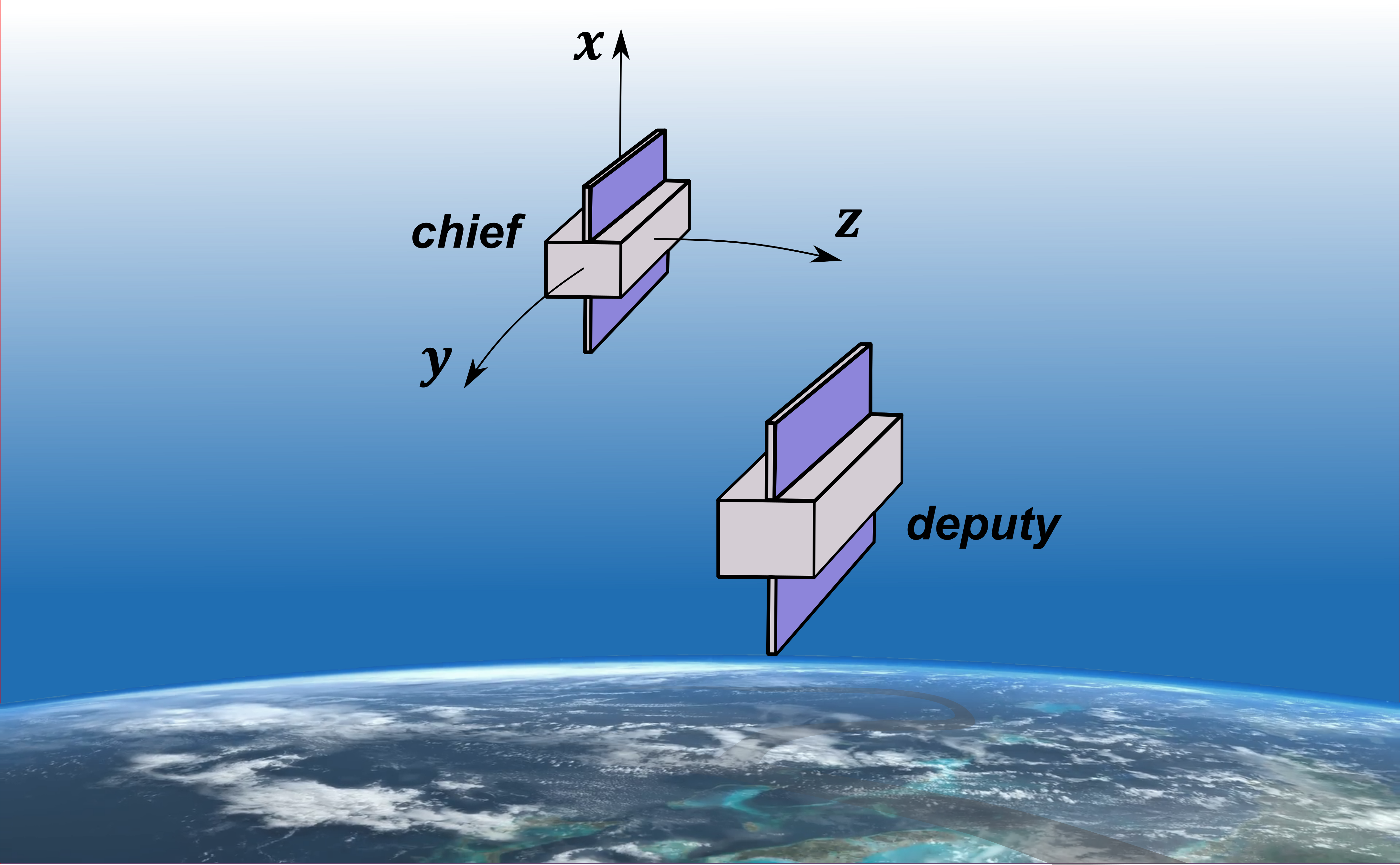}
	\caption{Graphical visualization of the satellite geometries including the spherical LVLH frame centered at the chief.}
	\label{fig:Visualization}
\end{figure}

To simplify and generalize, throughout this article formation flight maneuvers of formations consisting of two satellites are considered. In any case, close proximity operations are avoided as the inherent risk of collision due to uncertainties and dynamic variations in the control force renders such maneuvers impractical. The satellites under investigation, which are of identical shape and size\footnote{A common strategy to facilitate the formation maintenance task.}, are 3U CubSats augmented with two additional external panels each. The panels, which can be solar panels, are oriented such that the control authority of both differential forces (lift and drag) is increased (see Fig. \ref{fig:Visualization}). The simple satellite design ensures that multiple reflections or shadowing effects, both of which cannot be taken into account using the panel method, are avoided. Following the CubeSat standard, the satellite bodies consist of 30 x 10 x 10 cm (length/width/height) cuboids and the panels, which are attached directly to the body, have a size of 30 x 0.5 x 12.5 cm each. A visualization of the satellite geometries including the spherical LVLH frame centered at the chief, in which all relative trajectories are presented in the following, is depicted in Fig. \ref{fig:Visualization}. 

Each satellite has a total mass of $ m_{C,D} = 5 kg $ and neither of them is equipped with any thrusting device. The corresponding symmetric moments of inertia tensor $ \left[ \boldsymbol{I}_{sat}\right] $ expressed in the body fixed coordinate frame is:
\def\I{
	\begin{bmatrix}
		0.0152 & 0 & 0  \\
		0 & 0.0490 & 0  \\
		0 & 0 & 0.0412 \\
\end{bmatrix}}
\begin{equation}
	\left[ \boldsymbol{I}_{sat} \right] =\I \ kgm^{2}
\end{equation}
In terms of attitude control, the satellites are equipped with three Astrofein RW1 Type A reaction wheels which have a moment of inertia around their rotational axis of $ 694.5 \cdot 10^{-9} \ kgm² $, a nominal maximum torque of $ 23 \cdot 10^{-6} \ Nm $ and a maximum rotational velocity of 16,380 rpm\footnote{https://www.astrofein.com/2728/img/Dateien/ASTROFEIN\%20RW1.pdf}. The maximum torque of the reaction wheels is used as the limit $ T_{w,max} $ for the control variable in the maneuver planner (see Eq. \ref{eq:Constraints3}). In addition, the admissible yaw angle range is defined as $ \psi_{C,D} \in \left[-90^{\circ}, 90^{\circ} \right]  $ and the angular velocity of the satellites limited to $ \dot{\psi}_{C,D} \in \left[-0.1^{\circ}/s, 0.1^{\circ}/s \right]  $. Thereby, it is ensured that a complete switch of a satellite from a minimum to a maximum drag configuration (from $ 0^{\circ} $ to $ 90^{\circ} $) requires at least 15 minutes and unwanted effects like chattering or bang-bang type control switches are avoided. Constant and moderate solar and geomagnetic activities are assumed and the values according to ISO 14222 are used \cite{ISO14222.}. The maneuver parameters which are valid for all subsequent cases are summarized in Tab. \ref{tab:ManeuverParameters}. Whereas the initial absolute states of the chief remain invariant and are therefore included in Tab. \ref{tab:ManeuverParameters}, the initial $ \boldsymbol{\rho}_{0} $ and desired final $ \boldsymbol{\rho}_{f} $ relative states of the deputy with respect to the chief vary with the respective maneuver cases and are therefore stated in the individual subchapter.
\begin{table}
	\begin{center}
		\footnotesize
		\begin{tabular}{c c c} 
			\hline
			Parameter & Unit & Value \\
			\hline \hline
			$ a_{C,0} $ & $ \left[ km \right]  $ & 6678.137 \\
			$ e_{C,0} $ & $ \left[ - \right]  $ & 0.001 \\
			$ i_{C,0} $ & $ \left[ ^{\circ} \right]  $ & 98 \\
			$ \Omega_{C,0} $ & $ \left[ ^{\circ} \right]  $ & 10 \\
			$ \omega_{C,0} $ & $ \left[ ^{\circ} \right]  $ & 30 \\
			$ \theta_{C,0} $ & $ \left[ ^{\circ} \right]  $ & 60 \\
			$ \psi_{C,D,0} $ & $ \left[ ^{\circ} \right]  $ & $ 0 $ \\
			$ \dot{\psi}_{C,D,0} $ & $ \left[ ^{\circ} \right]  $ & $ 0 $ \\
			$ \psi_{C,D,f} $ & $ \left[ ^{\circ} \right]  $ & $ 0 $ \\
			$ \dot{\psi}_{C,D,f} $ & $ \left[ ^{\circ} \right]  $ & $ 0 $ \\
			$ F_{10.7}= \bar{F}_{10.7}$ & sfu  & 140 \\
			$ A_{p} $ &  nT  & 15 \\
			$ t_{f,min} $ & $ \left[ h \right] $  & 23.4 \\
			$ t_{f,max} $ & $ \left[ h \right] $  & 24.9 \\
			$ t_{f,guess} $ & $ \left[ h \right] $  & 24.1 \\
			Epoch & - & \makecell{22/10/2016 \\ 00:00:00 (UTC)}  \\
			\hline
		\end{tabular}
		\caption{Relevant maneuver parameters which are valid for all subsequently discussed exemplary maneuver cases.}
		\label{tab:ManeuverParameters}
	\end{center}
\end{table}
For all maneuvers, the initial guess for the maneuver time $ t_{f,guess} $ is arbitrarily defined to $ n_{guess} = 16 $ orbital periods of the initial chief orbit $ T_{C,0} = 2 \pi \sqrt{\frac{a_{C,0}}{\mu_{e}}}$. This results in $ t_{f,guess} = 24.1 \ h$. The limits are set to $ n_{min} = n_{guess} -\frac{1}{2} $ and $ n_{max} = n_{guess} +\frac{1}{2} $ which results in $ t_{f,min} =  23.4 \ h $ and  $ t_{f,max} = 24.9 \ h $.\\

Simulations are performed using an Intel® CoreTM i7 - 7700HQ CPU @ 2.80GHz and MATLAB® R2018a. The settings of the GPOPS-II software employed to schedule the optimal maneuver trajectory are listed in Tab. \ref{tab:GPOPSParameter}. 
\begin{table}
	\begin{center}
		\footnotesize
		\begin{tabular}{c c} 
			\hline
			 Max. iterations & 4 \\
			 Method & Hp-LiuRao \\
			 Tolerance & 3e-5 \\
			 Colpointsmin & 6 \\
			 Colpointsmax & 20 \\
			 R & 1.6\\
			 IPOPT tolerance & 1e-10 \\
			 Derivatives supplier & SparseFD \\
			 Derivatives level & First \\
			 Derivatives dependencies & sparseNaN \\
			 Derivatives stepsize & 1e-10 \\
			 Scales numsamples & 200\\
			\hline
		\end{tabular}
		\caption{Settings of the GPOPS-II software employed to schedule the optimal maneuver trajectory.}
		\label{tab:GPOPSParameter}
	\end{center}
\end{table}

\subsection{Pre-processing results}
\label{ssec:sPreProRes}

\subsubsection{Analytic density model fit}
\begin{figure}
		\centering
	\includegraphics{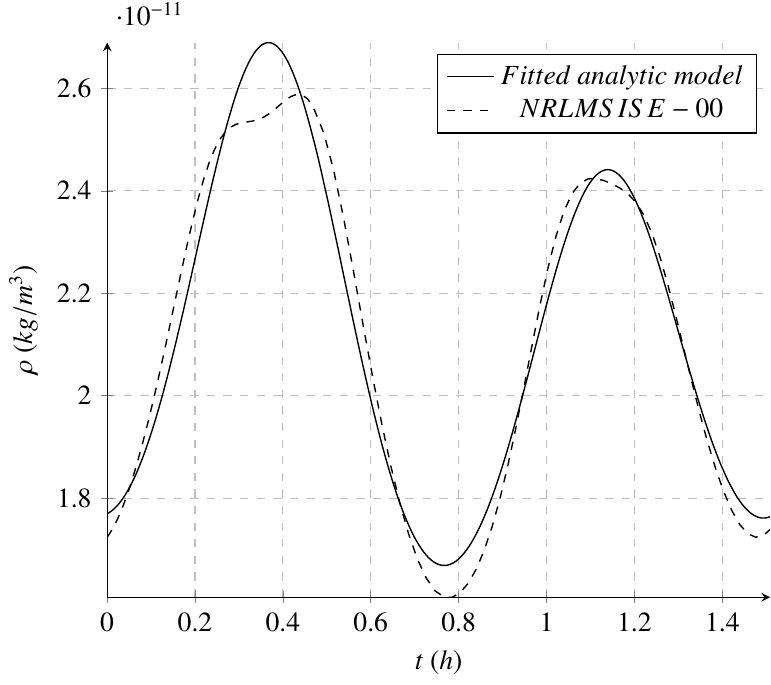}
	\caption{Comparison of the output of the fitted analytic density model and the reference data, which is produced using 	the NRLMSISE-00 environmental model \cite{Picone.2002}.}
	\label{fig:DensityFit}
\end{figure}

In the first of the two performed pre-processing steps, the four coefficients ($ A $, $ B $, $ C $, $ D $) of the analytic density model from Eq. \ref{eq:DenstyModel} are fitted in a sense that the mean squared error between the model results and the density values predicted by the NRLMSISE-00 environmental model \cite{Picone.2002} is minimized (see Eq. \ref{eq:MSE}). Fig. \ref{fig:DensityFit} shows a comparison between the density values predicted by the fitted analytic density model as well as the reference data calculated via the NRLMSISE-00 model \cite{Picone.2002} over one orbital period of the chief $ T_{C,0}$. Whereas some mismatches in the predicted density values are inevitable, the model is able to depict the most relevant dynamic variations occurring during the orbital revolution with a significantly reduced computational burden as no additional conversions are required.

\subsubsection{Fitting of the aerodynamic properties of the spacecraft}
\begin{figure*}
	\setcounter{subfigure}{0}
	\centering
	\subfloat[]{
	\includegraphics{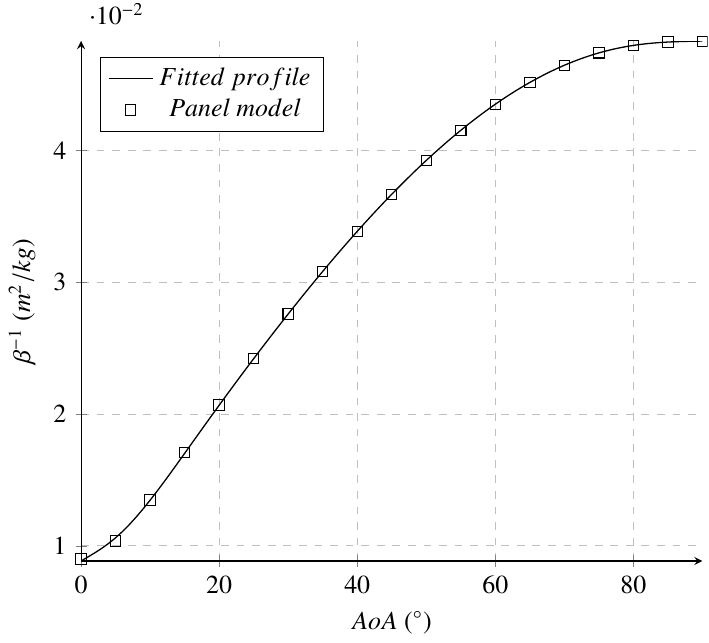}}
	\subfloat[]{
	\includegraphics{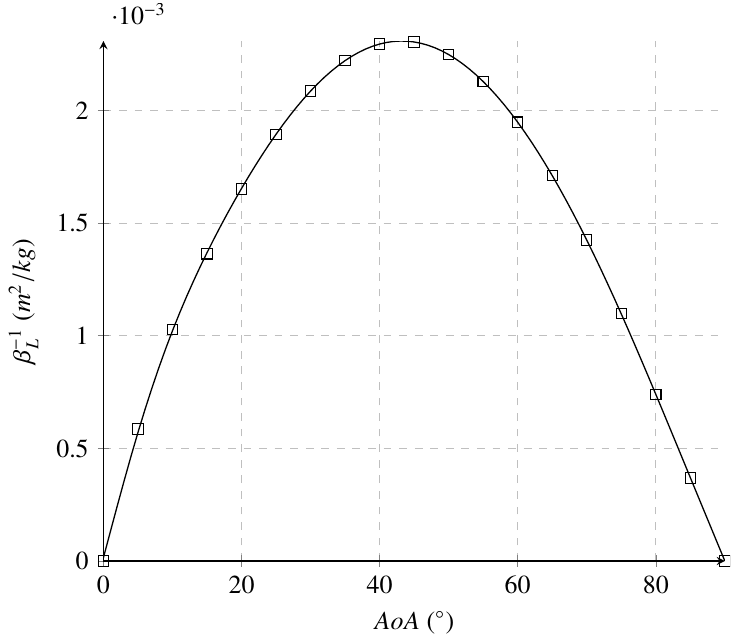}}
	\caption{Comparison of the fitted and reference ballistic coefficient $ \beta $ (left) and ballistic lift coefficient $ \beta_{L} $ (right) of the chief spacecraft. The reference data is calculated using the panel model with Sentman's GSI model \cite{Sentman.1961}.}
	\label{fig:BCFit}
\end{figure*}

In the second pre-processing step, the aerodynamic properties of the spacecraft, represented by $ \beta = f(AoA) $ and $ \beta_{L} = f(AoA) $, are fitted to the reference data calculated via the panel method applied with Sentman’s GSI model (see subchapter 3.1.2) for an angle-of-attack range of $ AoA \in \left[0^{\circ}, 90^{\circ} \right]  $. The results of this fit as well as the reference data are plotted in Fig. \ref{fig:BCFit}. Notably, the values are plotted reciprocally ($ \beta^{-1} $ and $ \beta_{L}^{-1} $) as this is a more vivid depiction of the aerodynamic sensitivity: an increase in $ \beta^{-1} $ indicates a higher sensitivity towards the aerodynamic drag force and, in analogy, an increase in $ \beta_{L}^{-1} $ indicates a higher sensitivity towards the aerodynamic lift force. Due to the symmetrical shape of the spacecraft, a similar profile results for a rotation in the opposite direction. 

From Fig. \ref{fig:BCFit}, the most important aerodynamic characteristics of the spacecraft under investigation can be interfered: the spacecraft are orbiting in their minimal drag configuration (minimal sensitivity towards the aerodynamic drag force) for an angle-of-attack of $ AoA_{min}=0^{\circ} $ and the resulting drag increases monotonously up to a maximum at the upper limit of $ AoA_{max}= \pm 90^{\circ} $. Vice versa, no aerodynamic lift is produced at the extreme values of the analysed AoA range ($ AoA_{min}$ and $ AoA_{max}$) but a maximum occurs at a value of around $ AoA_{lift,max} \approx \pm 45^{\circ}$. In any case, the values for $ \beta_{L}^{-1} $ and therefore the achievable lift forces are significantly lower than for $ \beta^{-1} $, a circumstance which matches the in-orbit observations. With respect to the methodology of differential lift and drag, these findings give insights into the fundamental nature of simultaneous in- and out-of-plane control via satellite aerodynamics: the AoA combination for which maximum differential drag values are achieved is if one satellite is orbiting at $ AoA_{min}$ and the second one at $ AoA_{max}$. For this combination, however, no differential lift is created at all. In contrast to this, maximum differential lift values can be achieved if both satellites counter rotate to AoA values of $ AoA_{lift,max} \approx \pm 45^{\circ} $ for which, however, no differential drag is created. Thus, in order to simultaneous create differential lift and drag, angle-of-attack combinations which differ from the just discussed extreme cases are required. In order to simultaneously minimize the resulting decay, both satellites aim at best possibly orbiting at the minimum possible AoA values required to fulfill the respective maneuver task.

\subsection{Maneuver results}
\label{ssec:ManRes}
In this subchapter, the maneuver results for three different maneuver cases of fundamental difference are presented. Thereby, the vast range of applicability and powerfulness of the developed approach shall be indicated. To increase the vividness of the desired maneuvers, the inital and final relative states of the deputy with respect to the chief are expressed in terms of the parameters $ \rho $, $ d $, $ \rho_{z} $, $ v_{d} $, $ \tilde{\alpha}_{0} $, and $ \tilde{\beta}_{0} $ introduced in subchapter \ref{ssec:CW}. The resulting maneuver trajectories are displayed in relative Cartesian states $ \boldsymbol{\rho} $, which are calculated from the nearly-nonsingular elements $ \delta \boldsymbol{\bar{\mathcal{E}}}_{ns} $ and $ \boldsymbol{\bar{\mathcal{E}}}_{ns,C} $ via the mapping presented by Schaub et al. \cite{Schaub.2018,Schaub.2004}, which is listed in \ref{Conversion} for reference. 

In each case, the main results of the maneuver, that is the final maneuver time $ t_{f} $, the resulting orbital decay of the chief:
\begin{equation}
	\Delta  a_{C} = a_{C,0}-a_{C,f}
\end{equation}
and the time required to find the optimal solution for the maneuver $ t_{plan} $ are summarized in a respective table.

\subsubsection{Case 1: In-plane formation re-phasing maneuver with simultaneous out-of-plane regulation}
\label{sssec:Case1}
In the first maneuver example, case 1, a re-phasing maneuver of an in-plane formation during which the order of the leader / follower configuration is switched is conducted. Simultaneously, a residual initial out-of-plane motion $ \rho_{z,0} \neq 0$ shall be regulated. The respective initial and final states of the deputy with respect to the chief are summarized in Tab. \ref{tab:Case1Parameter}. 

\begin{figure}
	\centering
	\includegraphics{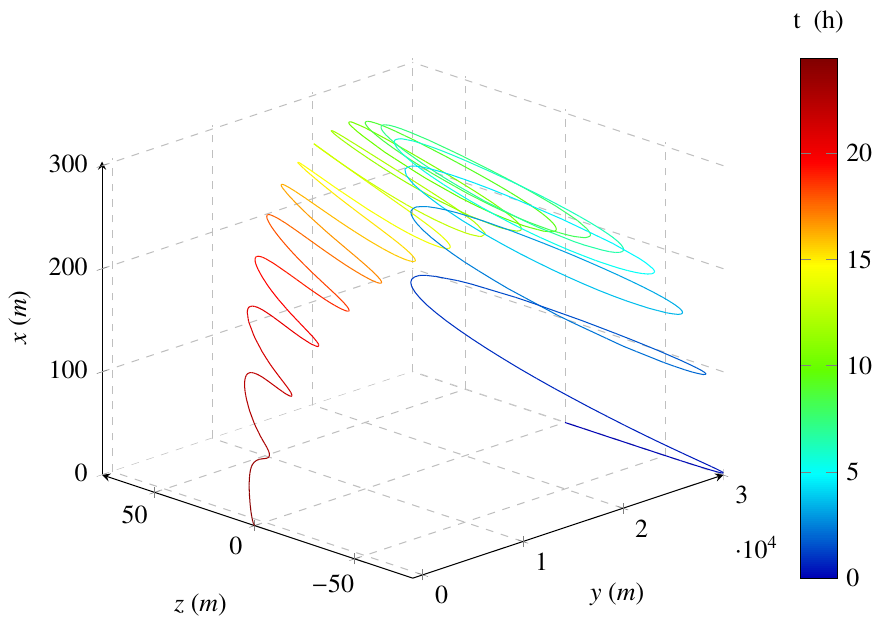}
	\caption{Relative maneuver trajectory of the deputy with respect to the chief plotted in the LVLH frame centred at the chief for an in-plane formation re-phasing maneuver with simultaneous out-of-plane regulation.}
	\label{fig:Case13D}
\end{figure}

\begin{figure}
	\centering
	\includegraphics{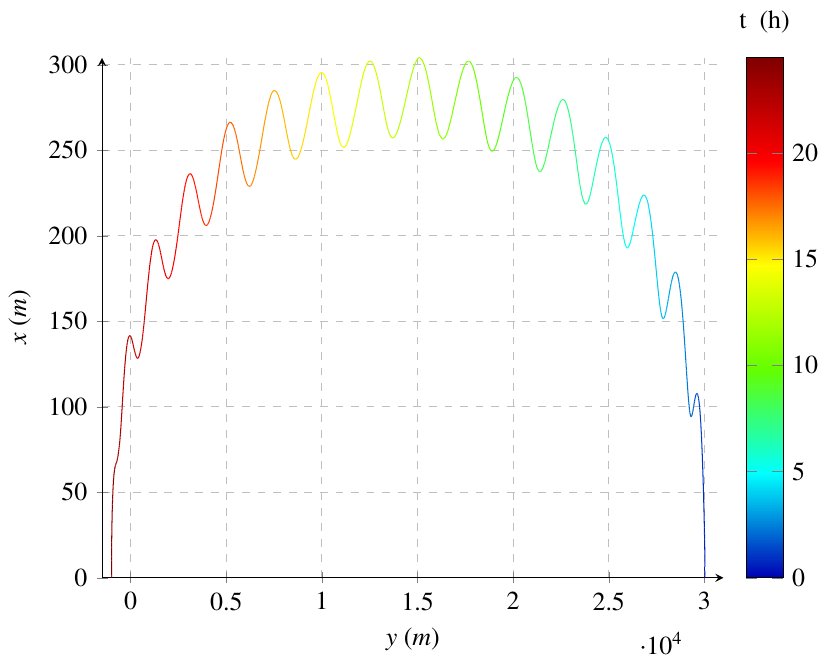}
	\caption{Relative in-plane trajectory of the deputy with respect to the chief plotted in the LVLH frame centred at the chief for an in-plane formation re-phasing maneuver with simultaneous out-of-plane regulation.}
	\label{fig:Case12D}
\end{figure}

\begin{figure*}
	\setcounter{subfigure}{0}
	\centering
	\subfloat[]{
	\includegraphics{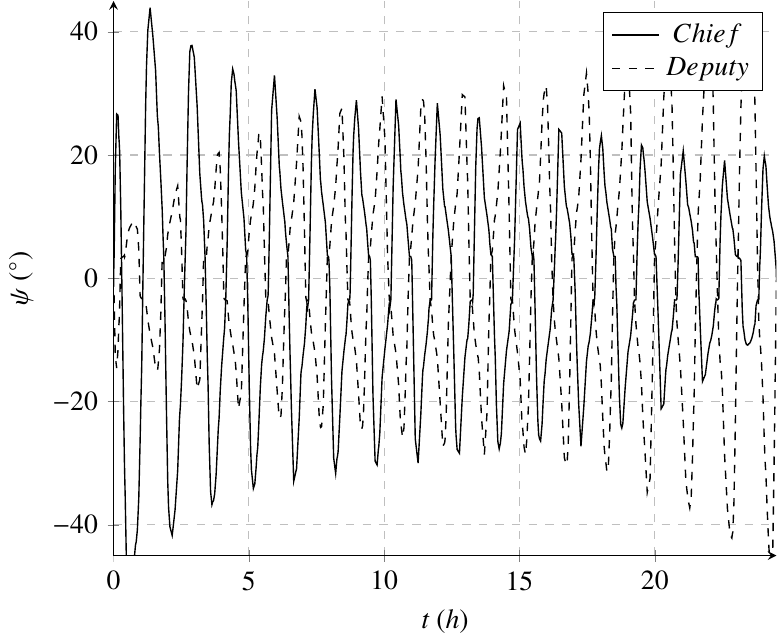}}
	\subfloat[]{
	\includegraphics{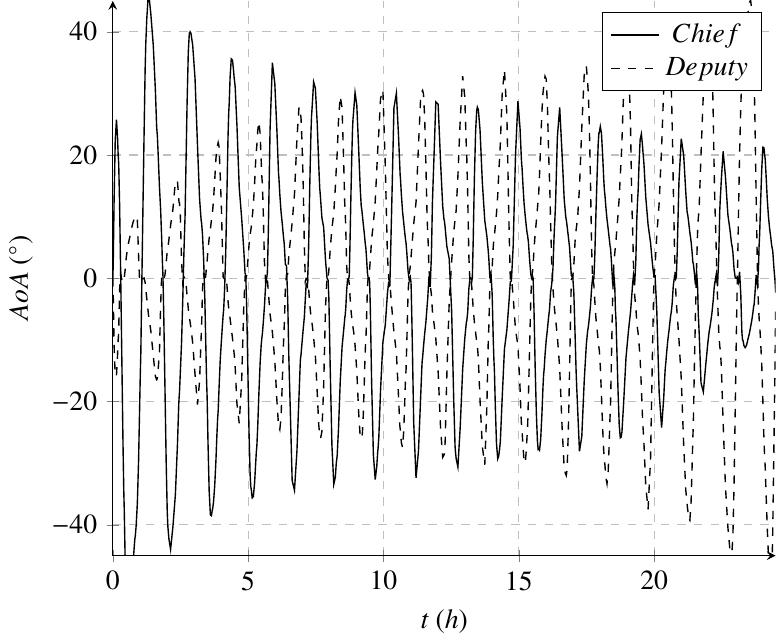}}
\caption{Resulting yaw angle $ \psi(t) $ (left, a)) and angle-of-attack $ AoA(t) $ (right, b)) profile of both satellites for an in-plane formation re-phasing maneuver with simultaneous out-of-plane regulation.}
	\label{fig:Case1Angles}
\end{figure*}

The resulting three-dimensional relative maneuver trajectory of the deputy with respect to the chief in the LVLH frame of the chief is plotted in Fig. \ref{fig:Case13D}, where a color bar was added to indicate the maneuver time. In addition, the in-plane maneuver trajectory is shown in Fig. \ref{fig:Case12D} and the corresponding yaw angle $ \psi(t) $ and AoA profiles $ AoA(t) $ for both satellites is depicted in Fig. \ref{fig:Case1Angles}. The main results of the maneuver are summarized in Tab. \ref{tab:Case1Results}.

\begin{table}
	\begin{center}
		\footnotesize
		\begin{tabular}{c | c c c c c c} 
			\hline
			 & $ v_{t} \left[ m/s^{2} \right] $  & $ \rho \left[ m \right] $ & $ \alpha_{0} \left[ ^{\circ} \right]  $  & $ \rho_{z} \left[ m \right] $ & $ \beta_{0} \left[ ^{\circ} \right]  $ & $ d \left[ km \right]  $ \\
			\hline \hline
			$ t_{0} $ & 0 & 0 & 0 & 80 & 90 & 30 \\
			$ t_{f} $ & 0 & 0 & 0 & 0 & 0 & -1 \\
			\hline
		\end{tabular}
		\caption{Initial and final formation configuration for an in-plane formation re-phasing maneuver with simultaneous out-of-plane regulation (case 1).}
		\label{tab:Case1Parameter}
	\end{center}
\end{table}
\begin{table}
	\begin{center}
		\footnotesize
		\begin{tabular}{c c c} 
			\hline
			$ t_{f} \left[ h \right]$ & $ \Delta  a_{C} \left[ m \right] $ & $ t_{plan} \left[ h \right]$  \\
			 24.49 & 1717.32 & 3.62  \\
			\hline
		\end{tabular}
		\caption{Main results of the in-plane formation re-phasing maneuver with simultaneous out-of-plane regulation (case 1).}
		\label{tab:Case1Results}
	\end{center}
\end{table}

\subsubsection{Case 2: In-plane formation into centered, bounded relative motion maneuver}
\label{sssec:Case2}
In the second maneuver example, case 2, the satellites shall be guided from an in-plane formation into a centered, bounded relative motion ellipse during the maneuver. The respective initial and final states of the deputy with respect to the chief are listed in Tab. \ref{tab:Case2Parameter}. 

\begin{figure}
	\centering
\includegraphics{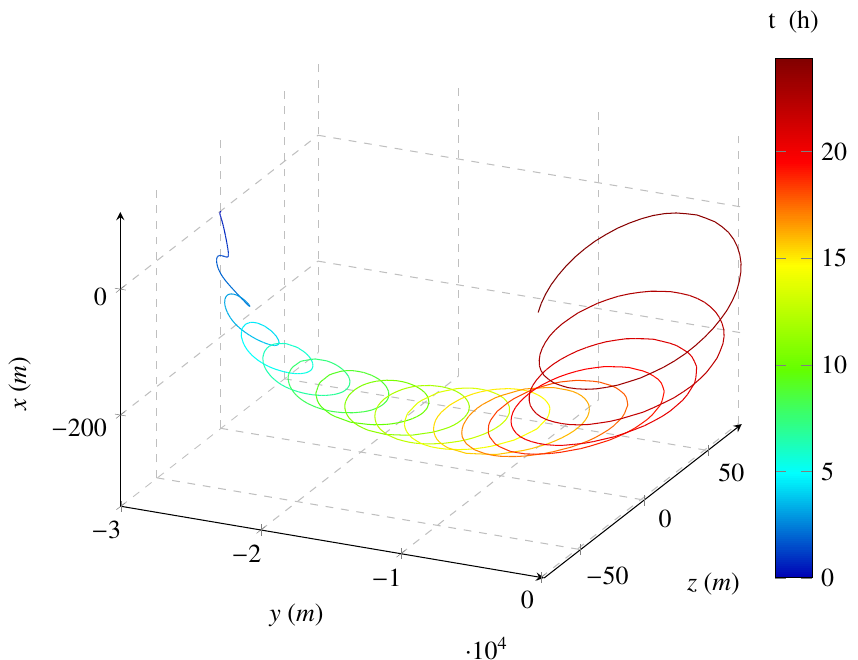}
	\caption{Relative maneuver trajectory of the deputy with respect to the chief plotted in the LVLH frame centred at the chief for an in-plane formation into a centred, bounded relative motion maneuver.}
	\label{fig:Case23D}
\end{figure}

\begin{figure}
	\centering
\includegraphics{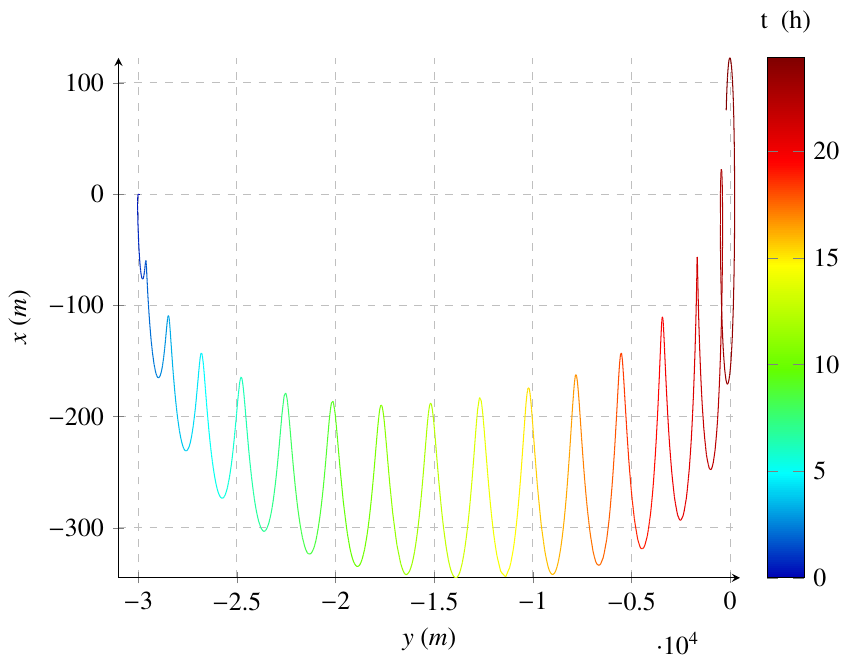}
	\caption{Relative in-plane trajectory of the deputy with respect to the chief plotted in the LVLH frame centred at the chief for an in-plane formation into a centred, bounded relative motion maneuver.}
	\label{fig:Case22D}
\end{figure}

\begin{figure*}
	\setcounter{subfigure}{0}
	\centering
	\subfloat[]{
\includegraphics{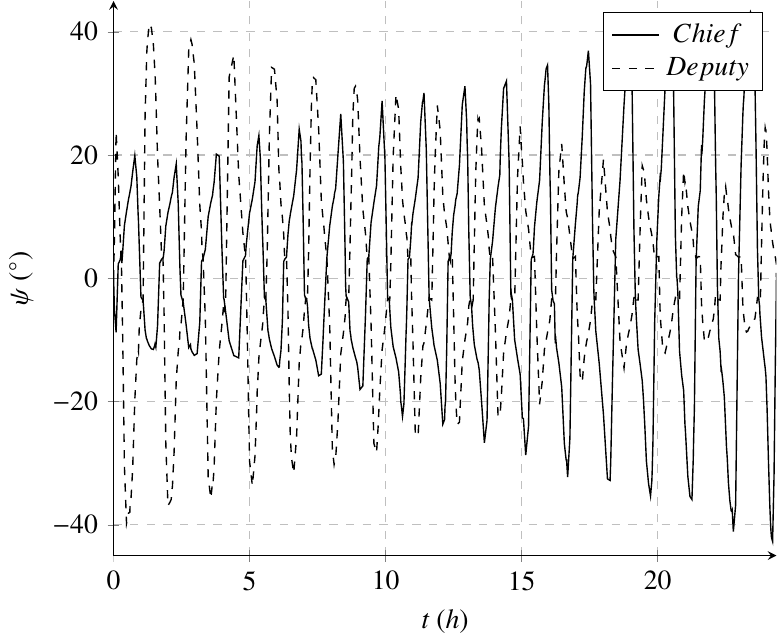}}
	\subfloat[]{
\includegraphics{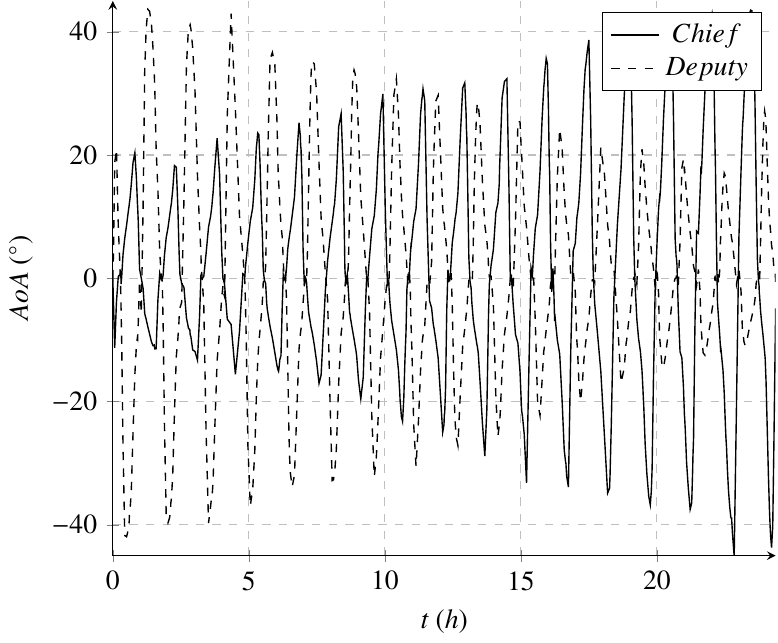}}
	\caption{Resulting yaw angle $ \psi(t) $ (left, a)) and angle-of-attack $ AoA(t) $ (right, b)) profile of both satellites for an in-plane formation into a centred, bounded relative motion maneuver.}
	\label{fig:Case2Angles}
\end{figure*}

The resulting three-dimensional relative maneuver trajectories of the deputy with respect to the chief plotted in the LVLH frame centred at the chief are depicted in Fig. \ref{fig:Case23D}, the in-plane relative motion trajectory in Fig. \ref{fig:Case22D} and the corresponding yaw angle $ \psi(t) $ and angle-of-attack $ AoA(t) $ profiles for both satellites are depicted in Fig. \ref{fig:Case2Angles}. The main results of the maneuver are summarized in Tab. \ref{tab:Case2Results}.\\

\begin{table}
	\begin{center}
		\footnotesize
		\begin{tabular}{c | c c c c c c} 
			\hline
			& $ v_{t} \left[ m/s^{2} \right] $  & $ \rho \left[ m \right] $ & $ \alpha_{0} \left[ ^{\circ} \right]  $  & $ \rho_{z} \left[ m \right] $ & $ \beta_{0} \left[ ^{\circ} \right]  $ & $ d \left[ km \right]  $ \\
			\hline \hline
			$ t_{0} $ & 0 & 0 & 0 & 0 & 0 & -30 \\
			$ t_{f} $ & 0 & 125 & 0 & 80 & 90 & 0 \\
			\hline
		\end{tabular}
		\caption{Initial and final formation configuration for an in-plane formation into a centered, bounded relative motion maneuver (case 2).}
		\label{tab:Case2Parameter}
	\end{center}
\end{table}
\begin{table}
	\begin{center}
		\footnotesize
		\begin{tabular}{c c c} 
			\hline
			$ t_{f} \left[ h \right]$ & $ \Delta  a_{C} \left[ m \right] $ & $ t_{plan} \left[ h \right]$  \\
			24.41  & 1669.47 & 3.04 \\
			\hline
		\end{tabular}
		\caption{Main results of the in-plane formation into a centered, bounded relative motion maneuver (case 2).}
		\label{tab:Case2Results}
	\end{center}
\end{table}

\subsubsection{Case 3: Centered, bounded relative motion variation maneuver}
\label{sssec:Case3}
In the third and final maneuver example, case 3, the out-of-plane motion of a centered, bounded relative motion ellipse shall be adjusted while the in-plane formation design shall remain unaltered. The respective initial and final states of the deputy with respect to the chief are listed in Tab. \ref{tab:Case3Parameter}.

\begin{figure}
	\centering
\includegraphics{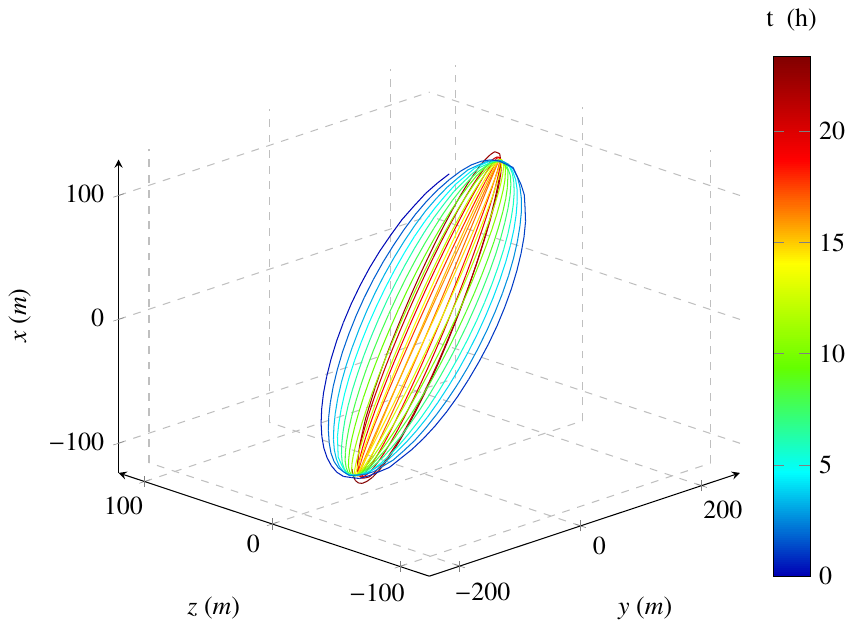}
	\caption{Relative maneuver trajectory of the deputy with respect to the chief plotted in the LVLH frame centred at the chief for a centered, bounded relative motion variation maneuver.}
	\label{fig:Case33D}
\end{figure}

\begin{figure}
	\centering
\includegraphics{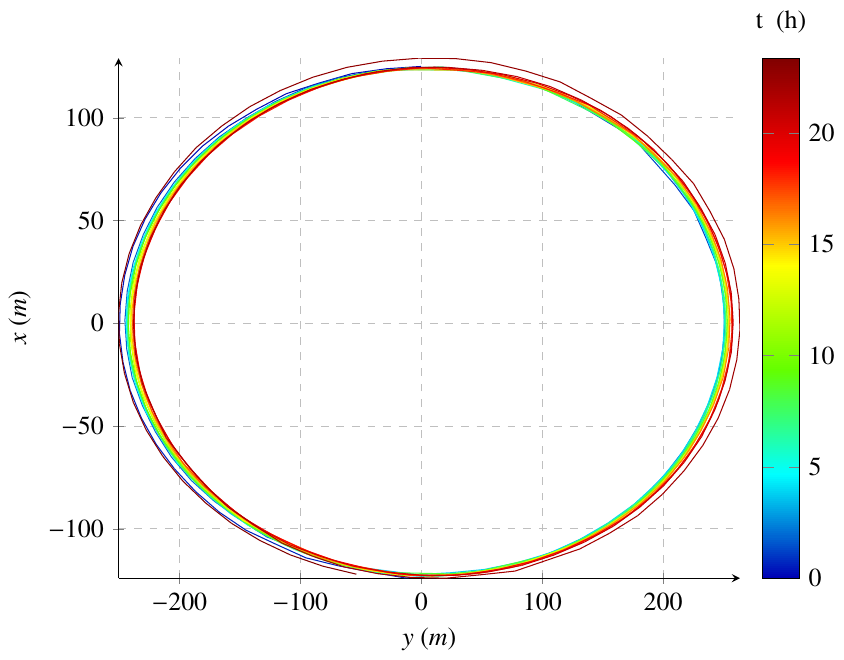}
	\caption{Relative in-plane trajectory of the deputy with respect to the chief plotted in the LVLH frame centred at the chief for a centered, bounded relative motion variation maneuver.}
	\label{fig:Case32D}
\end{figure}

\begin{figure*}
	\setcounter{subfigure}{0}
	\centering
	\subfloat[]{
\includegraphics{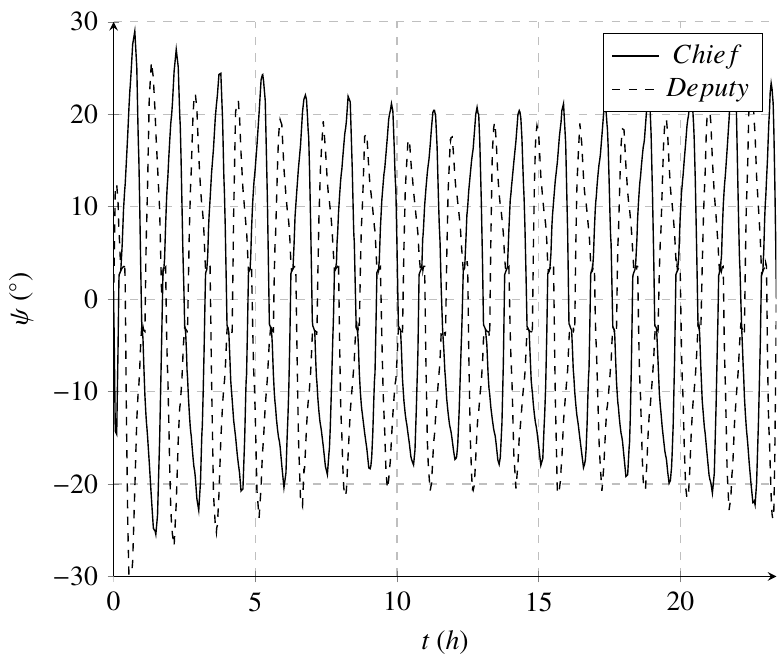}}
	\subfloat[]{
\includegraphics{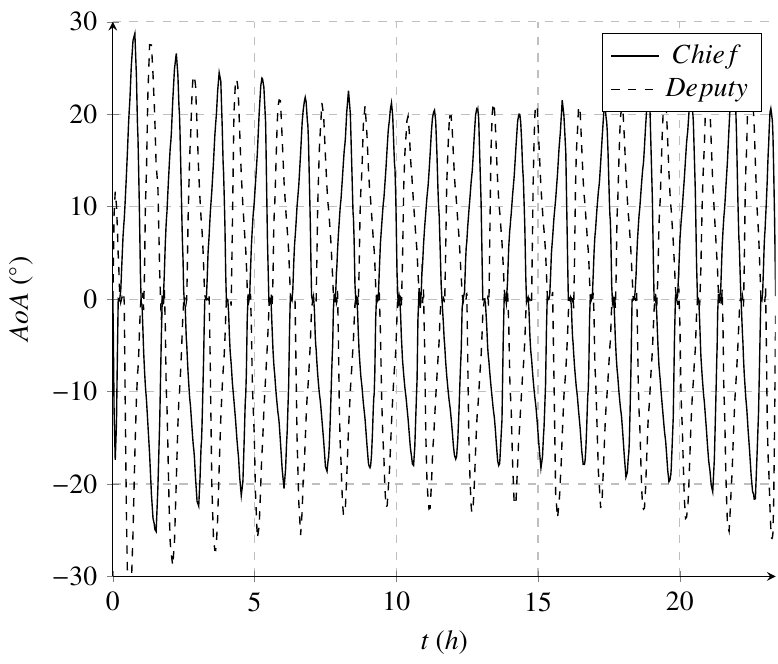}}
	\caption{Resulting yaw angle $ \psi(t) $ (left, a)) and angle-of-attack $ AoA(t) $ (right, b)) profile of both satellites for a centered, bounded relative motion variation maneuver.}
	\label{fig:Case3Angles}
\end{figure*}

 The resulting three-dimensional relative maneuver trajectory is plotted in Fig. \ref{fig:Case33D} and the in-plane relative motion trajectory in Fig. \ref{fig:Case32D}. The corresponding yaw angle $ \psi(t) $ and angle-of-attack $ AoA $ profiles for both satellites are depicted in Fig. \ref{fig:Case3Angles}. Again, the phase angles of the initial and final formation design are chosen such that the formation is passively safe (see discussion in subchapter \ref{ssec:CW}). The main results of the maneuver are summarized in Tab. \ref{tab:Case3Results}. 

\begin{table}
	\begin{center}
		\footnotesize
		\begin{tabular}{c | c c c c c c} 
			\hline
			& $ v_{t} \left[ m/s^{2} \right] $  & $ \rho \left[ m \right] $ & $ \alpha_{0} \left[ ^{\circ} \right]  $  & $ \rho_{z} \left[ m \right] $ & $ \beta_{0} \left[ ^{\circ} \right]  $ & $ d \left[ m \right]  $ \\
			\hline \hline
			$ t_{0} $ & 0 & 125 & 0 & 80 & 90 & 0 \\
			$ t_{f} $ & 0 & 125 & 0 & 120 & 90 & 0 \\
			\hline
		\end{tabular}
		\caption{Initial and final formation configuration for a centered, bounded relative motion variation maneuver (case 3).}
		\label{tab:Case3Parameter}
	\end{center}
\end{table}
\begin{table}
	\begin{center}
		\footnotesize
		\begin{tabular}{c c c} 
			\hline
			$ t_{f} \left[ h \right]$ & $ \Delta  a_{C} \left[ m \right] $ & $ t_{plan} \left[ h \right]$  \\
			23.38  & 1372.79 & 5.10  \\
			\hline
		\end{tabular}
		\caption{Main results of the centered, bounded relative motion variation maneuver (case 3).}
		\label{tab:Case3Results}
	\end{center}
\end{table}

\subsubsection{Classification of the results}
\label{sssec:Classification}
The variety of the presented maneuvers allows to gain deep insight into the nature of simultaneous in- and out-of-plane relative motion control via yaw angle deviations:

\textbf{Case 1}: Fig. \ref{fig:Case1Angles} depicts vividly how the simultaneous in- and out-of-plane control via yaw angle deviations is accomplished: in the initial formation design, the deputy is leading the in-plane formation by $ d_{0} = 30 km $. As during the maneuver the order shall be switched ($ d_{f} = -1 km $), the chief has to catch up with respect to the deputy during the maneuver. To do so, the chief is guided into a lower orbit with a correspondingly shorter orbital period. This is achieved by increasing its surface area perpendicular to the flow via increased angles-of-attack. As a consequence, the rate of decay of the chief is significantly larger then the one of the deputy ($ \delta \dot{a}< 0 $) and $ \delta a $ decreases accordingly. As during the maneuver also the residual out-of-plane motion shall be regulated, both satellites vary their angles-of-attack in a oscillating, coordinated and counteracting fashion. Thereby, following the nature of the out-of-plane satellite relative motion, the sign of the yaw angles switches twice per orbit. At this point, it must be emphasized again that it is not the yaw angle $ \psi  $ but the angle-of-attack $ AoA $ which ultimately governs the control forces in case of aerodynamically controlled satellite formation flight. This is depicted vividly in Fig. \ref{fig:Case1Angles}. As the maneuver progresses, the absolute values of the angle-of-attack of the chief gradually decrease, whereas, vice versa, the respective values for the deputy increase. This is required as after the maneuver, a stable formation shall result and consequently the semi-major axes of both spacecraft have to be adjusted respectively. 

\textbf{Case 2}: The maneuver assessed in case 2 differs from the maneuver analyzed in case 1 in two ways: 1) in this case, the order of the initial formation design is switched ($ d_{0} = -30 km $), meaning that the deputy is \textit{trailing} the in-plane formation and 2), that the final formation design consists of a passively stable, three-dimensional relative motion ellipse. Accordingly, the angle-of-attack profile $ AoA(t) $ depicted in Fig. \ref{fig:Case2Angles} is opposed to the profile depicted in Fig. \ref{fig:Case1Angles}. However, since the maneuver distance in the in-plane direction, the desired out-of-plane adjustments as well as the available maneuver time is within a comparable range to case 1, the absolute values of the $ AoA(t) $ correspond to the values from maneuver case 1. After the maneuver, a stable formation is ensured again by adjusting the semi-major axes of both spacecraft. Notably, by choosing the phase angles of the final formation design such that $ \beta_{0}= \alpha_{0} + \frac{\pi}{2} $, a passively safe formation is established as the trajectory never crosses the along-track axis (see discussion in subchapter \ref{ssec:CW}). 

\textbf{Case 3}: This maneuver substantially differs from the previously discussed maneuver examples (case 1 and 2) as the center of bounded relative motion ellipse of the deputy before and after the maneuver is located at the chief. To achieve an adjustment of the out-of-plane motion while maintaining the in-plane formation design, only marginal differences in the drag magnitudes but large differences in the lift magnitudes are required. This can be achieved when both satellites counter rotate  to angles-of-attack of similar magnitude ($ AoA_{C} \approx -AoA_{D} $), which is vividly depicted in Fig. \ref{fig:Case3Angles}. This, however, is a rather 'expensive' maneuver in terms of the resulting orbital decay as no adjustments to the in-plane motions are made. Nevertheless, the planning tool is able to schedule an optimal maneuver in this case, too.\\

\textbf{Planning efficient maneuver sequences}: The insights gained from the presented maneuver cases can be used to briefly touch on the planning of most efficient future maneuver sequences: for pending in-plane adjustment maneuvers, e.g. in-plane formation reconfiguration maneuvers, applying yaw angle deviations instead of pitch angle deviations allows to simultaneously adjust the out-of-plane motion basically at no cost (in terms of resulting decay). This out-of-plane control could also be used for maintenance purposes, i.e. to correct for perturbing accelerations caused by $ J_{2} $. A two-phased maneuver sequence, however, in which the in-plane motion is controlled first and, in a subsequent second step, the out-of-plane motion is adjusted accordingly (or vice versa), would result in significantly higher levels of decay. Notably, if for any other reason a two-phase maneuver should be more suitable, the tool is nevertheless able to schedule the optimal sequence.\\

Getting such insights, which are critical for the planning of most efficient future maneuver sequences, would be at least challenging if it weren't for a dedicated and high fidelity planning tool. 

\subsection{Summary and discussion}
\label{ssec:SumDis}

\subsubsection{Summary}
In this article, a planning tool for three-dimensional satellite formation flight maneuvers via differential aerodynamic forces has been presented. The tool allows to schedule any desired maneuver trajectory within the bounds of possibilities. In summary, the major assets of tool are as follows:
\begin{enumerate}
	\item Differential lift and drag can be applied \textit{simultaneously} via deviations of the yaw angles $ \psi $ of the two satellites so that all three translational degrees of freedom become be controllable;
	\item The maneuver is planned optimal in a sense the decay during the maneuver is minimized. Thereby, the practicability and sustainability of the methodology is substantially increased.
	\item The planning tool is able to take all dominant perturbing effects in VLEO, i.e. the $ J_{2} $ effect and aerodynamic forces, into account. In terms of aerodynamic forces, varying atmospheric densities, the co-rotation of the Earth as well as the fundamental natures of gas-surface interactions, are considered;
	\item Arbitrary relevant formation flight maneuvers within the bounds of possibilities can be planned and the desired initial and final relative states defined vivid and well known formation parameters;
	\item The resulting yaw angle profile avoids chattering or bang-bang type control switches so that it can be easily tracked by the attitude control systems.
\end{enumerate}
All respective boundary conditions can be varied highly flexible so that the developed methodology can now be applied to perform parameter studies aiming to explore and outline the design space of possible maneuver variants. The boundary conditions which can be defined by the user comprise:
\begin{enumerate}
	\item The maneuver epoch;
	\item The initial absolute mean states of the chief;
	\item The initial $ \boldsymbol{\rho}_{0} $ and final $ \boldsymbol{\rho}_{f} $ relative states of the deputy;
	\item The geometries and masses of the respective satellites;
	\item The GSI model which is applied in the panel method;
	\item The value for the energy accommodation coefficient $ \alpha_{T} $ (or whether the SESAM model shall be applied);
	\item The solar and geomagnetic activity proxies $ F_{10.7} $ and $ A_{p} $;
	\item The maximum rotational velocity of the satellites $ \dot{\psi}_{C,D} $ and the maximum torque that can be commanded to the reaction wheels $ T_{w,max} $.
\end{enumerate}
Over all, to the best of the authors’ knowledge, a tool with the given capabilities has not yet been presented in literature.

\subsubsection{Could superior control performance be gained by controlling roll, pitch and yaw? – A brief discussion}
\label{ssec:SupControl}
As of today, the proposed control strategies in terms of differential aerodynamic force creation via a rotation of asymmetrically shaped satellites comprise the creation of differential drag for the in-plane relative motion control via variations in the pitch angles $ \theta_{C,D} $ or, as discussed within this article, the simultaneous exploitation of differential lift and drag via yaw angle deviations $ \psi_{C,D} $. So far, however, a simultaneous control of two out of the three not to speak of all Euler angles has not been considered. Therefore, the question arises whether the simultaneous consideration of more than one Euler angle could possibly lead to superior maneuver sequences. Whereas an in-depth assessment based on dynamic simulations is left for future work, a brief discussion shall follow hereinafter.\\

The control of satellite formations comprises two sub-tasks, namely the control of the in- and the out-of-plane relative motion, which fundamentally differ. The in-plane relative motion is highly unstable and, due to the coupled nature of the two states, can be controlled by applying either differential lift (in the radial direction), differential drag (anti-parallel to the relative velocity vector) or any superposition of the two specific forces. However, since according to the VOP equations a perpendicular force components is significantly less effective in changing the orbit geometry compared to the along-track force component and moreover for state-of-the art satellite surface materials lift coefficients $ C_{L} $ experienced in orbit so far are significantly smaller than the respective drag coefficients $ C_{D} $, differential drag is currently unquestionable the control method of choice for the in-plane relative motion \cite{Walther.2020,Smith.2017,Smith.2019}. For the out-of-plane relative motion, however, the only suitable aerodynamic control option is to apply differential lift forces perpendicular to the orbital plane. Due to this reasoning, applying yaw angle deviations, as proposed in this article, is the best suited option to achieve three-dimensional relative motion control with only one control variable. Moreover, as the control authority of differential lift is very moderate anyways, any non-optimal application would render the methodology practically infeasible. However, this is not to say that in certain cases the control of more than one Euler angle and therefore the generation of differential forces in any possible direction could not lead to improved (or at least more flexible) maneuver sequences. To enable a simultaneous control of multiple Euler angles as well as a to subsequently perform in-depth assessments of possible advantages based on dynamic simulations is therefore considered part of our future work. Since for a problem to be solvable with GPOPS-II it need not to be linear, the presented planning tool provides a perfect basis for a future extension.

\section{Conclusion}
\label{sec:Conclusion}
In this article, a highly flexible tool to design optimal three-dimensional formation flight maneuvers which takes the most significant perturbing effects in VLEO, namely the $ J_{2} $ effect and atmospheric forces, into account is presented. During the planning process, varying atmospheric densities as well as the co-rotation of the atmosphere are considered. In addition to its flexible and high-fidelity nature, the major assets of the tool are that the in-and out-of-plane relative motion are controlled simultaneously via deviations in the yaw angles of the respective satellites and that the trajectory is planned optimal in a sense that the overall decay during the maneuver is minimized. Thereby, the remaining lifetime of the satellites after a maneuver is maximized and the practicability and sustainability of the methodology significantly increased. The vast range of applicability of the developed approach is indicated via three exemplary maneuver results which are of fundamental difference. In the near future, the methodology is foreseen to be applied to perform in-depth analysis which serves to explore and outline the design space of possible maneuver variants.

\subsection{Outlook}
\label{ssec:FutureWork}

\subsubsection{In-depth analysis}
The presented tool enables to plan optimal and three-dimensional satellite formation flight maneuvers via differential lift and drag with respect to a variety of input parameters. Due to its flexible approach, it can be applied for in-depth analyzes on the influences of different boundary conditions, such as the solar and geomagnetic activity, the satellite design (geometry and surface properties) or the respective maneuver task. The results obtained can be used to gain valuable insights into the methodology of differential lift and drag which will help to increase its practicability in a real mission scenario.

\subsubsection{Satellite design optimization}
With the presented tool, optimal maneuver trajectories for a given satellite geometry can be planned. Throughout the article, however, formation flight maneuvers of rather conventional 3U CubeSats augmented with two additional panels are investigated. Whereas this serves the generalization of the analysis, the satellite design parameters, i.e. shape and surface properties, are not at all optimized for their application or task. Therefore, it is aimed at identifying and developing optimal satellite designs with respect to their dedicated maneuver goal (cost function) such as e.g. being time optimal, minimizing the orbital decay values or achieving a best-possible trade-off. This would represent a first step towards including the satellite design in the optimization process and thus represent a more holistic approach.

\subsubsection{On-line compensation}
The presented maneuver tool schedules an open-loop control profile for an optimal maneuver trajectory taking the most relevant perturbing perfects in VLEO, namely the $ J_{2} $-effect and aerodynamic forces, into account. In reality though, un-modeled dynamics, e.g. higher harmonics of the Earth’s gravitational potential field, third body effects or solar radiation pressure, and uncertainties, predominantly in the atmospheric density and the aerodynamic coefficients, will inevitably cause the real trajectory to deviate from the scheduled path. Therefore, in analogy to the original approach \cite{DellElce.2015,DellElce.2015b}, a receding horizon model predictive control (MPC) approach is required to ensure a proper tracking of the scheduled trajectory. Developing a suitable compensator and verifying its effectiveness in a high-fidelity six degrees-of-freedom propagator is considered future work.

\subsubsection{Assessment of the performance of the planning tool}
Within the pre-processing stage, a set of complexity simplification strategies have been proposed to reduce the computational burden. So far, however, it has not it remains unclear how the accuracy and performance of the planning tool in the timespan of interest is impacted by the use of such strategies against a) using the original atmospheric model and b) using a the constant atmospheric density assumption. An in-depth assessment of this is foreseen in the near future.

\bibliographystyle{elsarticle-num} 
\bibliography{libraryPlanningTool}  

\appendix

\section{Gauss's Variational Equations for the nearly-nonsingular elements}
\label{GVE}
The nearly-nonsingular form of the GVE can be expressed as \cite{Roscoe.2015}:
\begin{equation}\label{eq:GVE1}
\frac{da}{dt}= \frac{2a^{2}}{h}\left[ \left( q_{1}\sin(u) - q_{2}\cos(u)\right) u_{x} + \frac{p}{r} u_{y}\right] 
\end{equation}
\begin{equation}\label{eq:GVE2}
	\begin{split}
			\frac{d \lambda}{dt}= \left[ \frac{-p}{h(1+\eta)}\left(q_{1}\cos(u) + q_{2}\sin(u)\right) - \frac{2\eta r}{h} \right] u_{x} \\
		+ \frac{p+r}{h(1+\eta)} \left( q_{1}\sin(u) - q_{2}\cos(u) \right) u_{y} \\
		- \frac{r \sin(u) \cos(i)}{h\sin(i)} u_{z} 
	\end{split}
\end{equation}
\begin{equation}\label{eq:GVE3}
	\frac{di}{dt} =\frac{r\cos(u)}{h} u_{z}
\end{equation}
\begin{equation}\label{eq:GVE4}
	\begin{split}
		\frac{dq_{1}}{dt} = \frac{p \sin(u)}{h}u_{x} + \frac{1}{h}\left[ \left( p+r\right)\cos(u)+rq_{1} \right]u_{y} \\
	+ \frac{r q_{2} \sin(u) \cos(i)}{h \sin(i)} u_{z}
	\end{split}
\end{equation}
\begin{equation}\label{eq:GVE5}
	\begin{split}
			\frac{dq_{2}}{dt} = \frac{-p \cos(u)}{h}u_{x} + \frac{1}{h}\left[ \left( p+r\right)\sin(u)+rq_{2} \right]u_{y} \\
		- \frac{r q_{1} \sin(u) \cos(i)}{h \sin(i)} u_{z}
	\end{split}
\end{equation}
\begin{equation}\label{eq:GVE6}
	\frac{d \Omega}{dt} = \frac{r \sin(u)}{h \sin(i)} u_{z}
\end{equation}
and the orbit equation in terms of nearly-nonsingular elements is \cite{Roscoe.2015}:
\begin{equation}\label{eq:GVE0}
	r = \frac{a\eta^{2}}{1+q_{1}\cos(u)+q_{2}\sin(u)}
\end{equation}
 
\section{Differential form of Lagrange's Planetary Equations}
\label{DLPE}
The Jacobian of the Lagrange's Planetary Equations is formed by populating the columns of $ \left[ \boldsymbol{A} \right]  $ with the partial derivatives with respect to each of the nearly-nonsingular mean elements \cite{Roscoe.2015}:
\begin{equation}\label{eq:LPE0}
\left[ \boldsymbol{A} \right] = \left[ a_{ij}\right] = \left[ \frac{\partial f_{i}}{\partial \bar{\boldsymbol{\mathcal{E}}}}_{ns}\right]  
\end{equation}
With the constant parameter $ \epsilon $ defined as in Schaub et al. \cite{Schaub.2000b}:
\begin{equation}\label{eq:LPE01}
\epsilon = J_{2} \left( \frac{R_{E}}{p}\right)^{2} n 
\end{equation}
the nonzero elements of $ \left[ \boldsymbol{A} \right]  $ are \cite{Roscoe.2015}:
\begin{equation}\label{eq:LPE1}
\frac{\partial f_{\lambda}}{\partial a} = \frac{-3n}{2a} - \frac{21 \epsilon}{8a}\left[\eta \left( 3 \cos(i)^{2} -1 \right) + \left(5 \cos(i)^{2} - 1 \right)  \right] 
\end{equation}

\begin{equation}\label{eq:LPE2}
	\frac{\partial f_{\lambda}}{\partial i} = \frac{-3 \epsilon}{4}\left( 3\eta+5\right) \sin(2i) 
\end{equation}

\begin{equation}\label{eq:LPE3}
	\frac{\partial f_{\lambda}}{\partial q_{1}} = \frac{3 \epsilon}{4\eta^{2}}\left[ 3\eta \left( 3 \cos(i)^{2} -1 \right)+4\left(5 \cos(i)^{2}-1 \right)  \right] q_{1} 
\end{equation}

\begin{equation}\label{eq:LPE4}
	\frac{\partial f_{\lambda}}{\partial q_{2}} = \frac{3 \epsilon}{4\eta^{2}}\left[ 3\eta \left( 3 \cos(i)^{2} -1 \right)+4\left(5 \cos(i)^{2}-1 \right)  \right] q_{2} 
\end{equation}

\begin{equation}\label{eq:LPE5}
	\frac{\partial f_{q_{1}}}{\partial a} = \frac{21 \epsilon}{8a}\left(5\cos(i)^{2}-1 \right) q_{2} 
\end{equation}

\begin{equation}\label{eq:LPE6}
	\frac{\partial f_{q_{1}}}{\partial i} = \frac{15 \epsilon}{4} q_{2} \sin(2i)
\end{equation}

\begin{equation}\label{eq:LPE7}
	\frac{\partial f_{q_{1}}}{\partial q_{1}} = \frac{-3 \epsilon}{\eta^{2}} \left(5 \cos(i)^{2} -1 \right) q_{1}q_{2}
\end{equation}

\begin{equation}\label{eq:LPE8}
	\frac{\partial f_{q_{1}}}{\partial q_{2}} = \frac{-3 \epsilon}{4} \left(1 + \frac{4q_{2}^{2}}{\eta^{2}} \right) \left(5 \cos(i)^{2} -1 \right) 
\end{equation}

\begin{equation}\label{eq:LPE9}
	\frac{\partial f_{q_{2}}}{\partial a} = -\frac{21 \epsilon}{8a}\left(5\cos(i)^{2}-1 \right) q_{1} 
\end{equation}

\begin{equation}\label{eq:LPE10}
	\frac{\partial f_{q_{2}}}{\partial i} = -\frac{15 \epsilon}{4} q_{1} \sin(2i)
\end{equation}

\begin{equation}\label{eq:LPE11}
	\frac{\partial f_{q_{2}}}{\partial q_{1}} = \frac{3 \epsilon}{4} \left(1 + \frac{4q_{1}^{2}}{\eta^{2}} \right) \left(5 \cos(i)^{2} -1 \right) 
\end{equation}

\begin{equation}\label{eq:LPE12}
	\frac{\partial f_{q_{2}}}{\partial q_{2}} = \frac{3 \epsilon}{\eta^{2}} \left(5 \cos(i)^{2} -1 \right) q_{1}q_{2}
\end{equation}

\begin{equation}\label{eq:LPE13}
	\frac{\partial f_{\Omega}}{\partial a} = \frac{21 \epsilon}{4a} \cos(i)
\end{equation}

\begin{equation}\label{eq:LPE14}
	\frac{\partial f_{\Omega}}{\partial i} = \frac{3 \epsilon}{2} \sin(i)
\end{equation}

\begin{equation}\label{eq:LPE15}
	\frac{\partial f_{\Omega}}{\partial q_{1}} = \frac{-6 \epsilon}{\eta^{2}} q_{1} \cos(i)
\end{equation}

\begin{equation}\label{eq:LPE16}
	\frac{\partial f_{\Omega}}{\partial q_{2}} = \frac{-6 \epsilon}{\eta^{2}} q_{2} \cos(i)
\end{equation}

\section{Nearly-nonsingular orbital element differences to relative orbit position conversion}
\label{Conversion}
Schaub et al. \cite{Schaub.2018,Schaub.2004} presented a direct mapping between orbit element differences $ \delta \boldsymbol{\bar{\mathcal{E}}}_{ns} $ and Cartesian states $ \boldsymbol{\rho} $ which assumes that the relative orbit radius $ \rho $ is small in comparison to the inertial chief orbit radius $ r $, is presented. Throughout this article, this mapping is used to calculate the relative maneuver trajectories. The relative position vector components $ \boldsymbol{\rho} =\left(x, \ y, \ z \right)^{T} $ are given in terms of orbit elements through:
\begin{equation}\label{eq:Maping1}
\begin{split}
x \approx \frac{r}{a}\delta a + \frac{V_{r}}{V_{t}} r \delta u -\frac{r}{p}\left(2aq_{1}+r\cos(u) \right) \delta q_{1} \\
- \frac{r}{p} \left(2aq_{2} +r \sin(u)\right) \delta q_{2}
\end{split}
\end{equation}

\begin{equation}\label{eq:Maping2}
	y \approx r\left(\delta u + \cos(i) \delta \Omega  \right) 
\end{equation}

\begin{equation}\label{eq:Maping3}
	z \approx r \left(\sin(u) \delta i - \cos(u)\sin(i)\delta \Omega \right) 
\end{equation}
where the chief radial and transverse velocity components $ V_{r} $ and $ V_{t} $ are defined as:
\begin{equation}\label{eq:Maping4}
	V_{r}=\dot{r}=\dfrac{h}{p}\left(q_{1}\sin(u)-q_{2}\cos(u) \right) 
\end{equation}

\begin{equation}\label{eq:Maping5}
		V_{t}=r\dot{u}=\dfrac{h}{p}\left(1+q_{1}\cos(u)+q_{2}\sin(u) \right) 
\end{equation}
\end{document}